\documentclass[a4paper,12pt]{article} 
\usepackage{epsfig} 
\usepackage{amssymb} 
\usepackage[tbtags]{amsmath}
\usepackage{upref} 
\usepackage{float}
\usepackage{wrapfig}
\usepackage{subfigure}
\usepackage{longtable}

\newcommand{\captions}{\sf\caption}

\def\tgb{\tan{\beta}}

\def\neumass{m_{\tilde\chi_1^0}}

\def\ufbiii{V_{UFB-3}}

\def\lsim{\raise0.3ex\hbox{$\;<$\kern-0.75em\raise-1.1ex\hbox{$\sim\;$}}}
\def\gsim{\raise0.3ex\hbox{$\;>$\kern-0.75em\raise-1.1ex\hbox{$\sim\;$}}}

\def\yzero{\smash{\hbox{$y\kern-4pt\raise1pt\hbox{${}^\circ$}$}}}

\def\s2{\frac{1}{\sqrt2}}

\def\beq{\begin{equation}} 
\def\eeq{\end{equation}} 
\def\beqa{\begin{eqnarray}} 
\def\eeqa{\end{eqnarray}}

\def\IF{\relax{\rm I\kern-.18em F}} 
\def\II{\relax{\rm I\kern-.18em I}} 
\def\IP{\relax{\rm I\kern-.18em P}} 
\def\IC{\relax\hbox{\kern.25em$\inbar\kern-.3em{\rm C}$}} 
\def\IR{\relax{\rm I\kern-.18em R}}

\def\Dsl{\,\raise.15ex\hbox{/}\mkern-13.5mu D} 
\def\IZ{Z\kern-.4em  Z} 
\def\bmat{\left(\begin{array}} 
\def\emat{\end{array}\right)} 
 
\def    \part          {\partial} 
\def    \be            {\begin{equation}} 
\def    \ee            {\end{equation}} 
\def    \bea           {\begin{eqnarray}} 
\def    \eea           {\end{eqnarray}}

\begin{document}

\pagestyle{empty}

\rightline{DESY 03-024}
\rightline{HIP-2002-57/TH}
\rightline{FTUAM 03/05}
\rightline{IFT-UAM/CSIC-03-09}
\rightline{hep-ph/0304115}
\rightline{April 2003}

\renewcommand{\thefootnote}{\fnsymbol{footnote}}
\vspace{1cm}
\begin{center}
{\large{\bf Neutralino-Nucleon Cross Section and \\Charge and Colour Breaking Constraints
\\[5mm]
}}
\vspace{0.5cm}
\mbox{
\sc{
D.G. Cerde\~no$^{1}$,
E. Gabrielli$^{2}$,
M.E. Gomez$^{3}$,
C. Mu\~noz$^{4}$
}
}
\vspace{1cm}

{\small
{\it 
$^1$ 
II. Institut f\"ur Theoretische Physik, Universit\"at Hamburg,\\
Luruper Chaussee 149, D-22761 Hamburg, Germany.\\
\vspace*{2mm}
\it $^2$ Helsinki Institute of Physics,
P.O. Box 64, FIN-00014 Helsinki, Finland.\\
\vspace*{2mm}
\it $^3$ 
Departamento de F\'{\i}sica and Grupo Te\'orico de  F\'{\i}sica de
Part\'{\i}culas, Instituto Superior T\'ecnico,  Av. Rovisco Pais, 1049-001
Lisboa, Portugal\\
\vspace*{2mm}
\it $^4$ Departamento de F\'{\i}sica
Te\'orica C-XI and Instituto de F\'{\i}sica
Te\'orica C-XVI,\\ 
Universidad Aut\'onoma de Madrid,
Cantoblanco, 28049 Madrid, Spain.
} 
}


\renewcommand{\thefootnote}{\alph{footnote}}

\vspace{1cm}

{\bf Abstract} 
\\[7mm]
\end{center}
\begin{center}
\begin{minipage}[h]{14.0cm}
We compute the neutralino-nucleon cross section in several 
supersymmetric scenarios, taking
into account all kind of constraints.
In particular, 
the constraints that the absence of dangerous charge
and colour breaking minima imposes on the parameter space
are studied in detail.
In addition, 
the most recent experimental constraints, such as
the lower bound on the Higgs mass,
the $b\to s\gamma$ branching ratio, and the
muon $g-2$ are considered.
The astrophysical bounds on the dark matter density
are also imposed on the 
theoretical computation of the relic neutralino density,
assuming thermal production.
This computation is relevant for the theoretical 
analysis of the direct detection of
dark matter in current experiments.
We consider first the supergravity scenario
with universal soft terms and GUT scale. In this scenario the
charge and colour breaking 
constraints turn out to be quite important, and
$\tan\beta\lsim 20$ is forbidden. Larger values of 
$\tan\beta$ can also be forbidden, depending on the value of 
the trilinear parameter $A$. 
Finally, we study supergravity
scenarios with an intermediate scale, and also with
non-universal scalar and gaugino masses where
the cross section can be very large.


\end{minipage}
\end{center}

\newpage

\setcounter{page}{1}
\pagestyle{plain}
\renewcommand{\thefootnote}{\arabic{footnote}}
\setcounter{footnote}{0}

\section{Introduction}

A weakly-interacting massive particle (WIMP) is one of the most
interesting candidates for the dark matter in the Universe.
Since WIMPs
would cluster gravitationally with ordinary stars in the galactic halos,
there is the hope of detecting relic 
WIMPs directly, by observing their elastic scattering on  
target nuclei through nuclear recoils \cite{contemporary}.
In fact, one of the current experiments, the DAMA collaboration, has
reported 
data favouring the existence of a 
WIMP signal \cite{experimento1}.
It was claimed that 
the preferred range of the WIMP-nucleon cross section
is 
$10^{-6}-10^{-5}$ pb 
for a WIMP mass between 30 and 270 GeV \cite{experimento1,halo}.
Unlike this spectacular result, other collaborations such as 
CDMS \cite{experimento2}, EDELWEISS \cite{edelweiss}, 
IGEX \cite{igex}, and ZEPLIN I \cite{zeplin1} 
claim to have excluded important regions of the DAMA 
parameter space.


In any case, due to these and other projected experiments, 
it seems very plausible that the dark matter 
will be found in the near future. 
In this situation, and assuming that the dark matter 
is a WIMP, it is natural to wonder how big 
the cross section for its direct detection can be.
The answer to this 
question depends on the particular WIMP considered.
The leading candidate in this class is the lightest 
neutralino \cite{contemporary}, $\tilde{\chi}^0_1$,
a particle 
predicted by the supersymmetric (SUSY) extension of the standard model (SM).


In this paper we will analyse the SUSY scenario
in the framework of supergravity (SUGRA). 
Working in this framework one makes several 
assumptions. In particular, the soft parameters,
i.e., gaugino masses,
scalar masses, and 
trilinear couplings, are generated once SUSY is broken through
gravitational interactions.
They are denoted at the grand unification 
scale ($M_{GUT} \approx 2\times 10^{16}$ GeV)
by $M_{a}$,
$m_{\alpha}$, and $A_{\alpha\beta\gamma}$,
respectively. 
Likewise, 
radiative electroweak symmetry breaking is imposed, and 
one loop renormalization group equations (RGEs) are
used to derive low-energy SUSY parameters.
Let us also recall that 
the Higgsino mass parameter $\mu$ 
is determined by the minimization of the Higgs effective 
potential. This implies 
\begin{equation}
\mu^2 = \frac{m_{H_d}^2 - m_{H_u}^2 \tan^2 \beta}{\tan^2 \beta -1 } - 
\frac{1}{2} M_Z^2\ ,
\label{electroweak}
\end{equation} 
where
$\tan\beta= \langle H_u^0\rangle/\langle H_d^0\rangle$ 
is the ratio of Higgs vacuum expectation values.
The effect of the one-loop corrections to the scalar potential
can be minimized by evaluating the $\mu$ term at the scale
$M_{SUSY}=\sqrt{m_{\tilde t_1} m_{\tilde t_2}}$ \cite{Nojiri,Gamberini}.


With these assumptions, the SUGRA framework still allows a large
number of free parameters. 
In order to have predictive power one usually assumes 
that the above soft parameters 
are 
universal at $M_{GUT}$.
This is the so-called minimal supergravity (mSUGRA) scenario,
where there are only four free parameters: 
$m$, $M$, $A$, and $\tan \beta$. In addition, the
sign of $\mu$ remains also undetermined.  
It is worth noticing that explicit string constructions
with these characteristics 
can be found \cite{dilaton}.
In any case, 
the general situation for 
the soft parameters in supergravity is to have a
non-universal structure \cite{dilaton}. 
For the case of the observable scalar masses
this is due to the non-universal couplings
in the K\"ahler potential
between the hidden sector fields breaking SUSY and the
observable sector fields.
For the case of the gaugino masses this is due to the
non-universality of the gauge kinetic functions associated to the
different gauge groups. Of course,
general string constructions, whose low-energy limit is SUGRA,
exhibit these properties \cite{dilaton}.


Although the cross section for the elastic scattering of
$\tilde{\chi}^0_1$
on nucleons in the mSUGRA and non-universal SUGRA frameworks  
has been examined exhaustively in the 
literature 
(for recent works see e.g. 
refs.~\cite{Ellis}-\cite{Birkedal}),
no analyses can be found concerning the constraints that
arise from imposing the absence of charge and colour breaking 
minima.
This is the aim of the present paper.

As is well known, the presence of scalar fields with colour and
electric
charge in SUSY theories induces the possible existence of 
dangerous charge and colour breaking minima, which would make
the standard vacuum unstable \cite{revmunoz}-\cite{emidio}.
The presence of these instabilities may imply either that the
corresponding model is inconsistent or that it requires non-trivial
cosmology to justify that the Universe eventually fell in the
phenomenologically realistic (but local) minimum \cite{revmunoz,cosmology}.
The constraints on the parameter space of the theory 
that arise from imposing 
the absence of these instabilities 
are very important \cite{clm1}, and
we will study their consequences on the 
neutralino-nucleon cross section.


The paper is organized as follows.
In Section~2 we will briefly review the potentially dangerous
charge and colour breaking directions in the field space of
SUGRA and how to avoid them.
On the other hand, a correct phenomenology is also essential
in our analysis of the neutralino-nucleon cross section, and we will discuss 
in Section~3 the most recent experimental and astrophysical 
constraints which can affect this computation.
Then, in the rest of the paper, we 
will re-evaluate the cross section taking into account
the charge and colour breaking constraints in addition to the
experimental and astrophysical ones. In particular, 
in Section~4
we will study the value of the cross section when
the mSUGRA scenario is considered.
First, we will analyse this scenario in the usual context
of a GUT scale,
$M_{GUT} \approx 2\times 10^{16}$ GeV,
and second we will discuss 
how these results are modified
when the GUT condition
is relaxed. In particular, we will consider the case of an 
intermediate scale. 
Finally, 
in Section~5, 
the most general situation in the context of SUGRA, 
non-universal scalar and gaugino masses, will
be studied.
These analyses have been carried out for different values of the
trilinear parameter $A$.
The conclusions are left for Section~6.



\section{Charge and colour breaking constraints}

The basic ingredient of our analysis concerns the constraints
associated with the existence of dangerous directions in the field space.
A complete analysis of this issue,
including
in a proper way the radiative corrections to the
scalar potential, was carried out in ref.~\cite{clm1}.
The most relevant results obtained there for our present task
are the following.

There are two types of constraints:
the ones arising from directions in the field-space along
which the (tree-level) potential can become unbounded from below (UFB),
and those arising from the existence of charge and color
breaking (CCB) minima in the potential deeper than the
standard minimum. By far, the most restrictive bounds are the
UFB ones, and therefore we will concentrate on them here.

There are three UFB directions, labelled as UFB-1, UFB-2, UFB-3
in \cite{clm1}. It is worth mentioning here that in general the
unboundedness is only true
at tree-level since radiative corrections eventually raise the potential for
large enough values of the fields, but still these minima can be deeper than
the realistic one (i.e. the SUSY SM vacuum) and thus dangerous.
The UFB-3 direction, which involves
the scalar fields
$\{H_u,\nu_{L_i},e_{L_j},e_{R_j}\}$ with $i \neq j$
and thus leads also to electric charge
breaking, yields the strongest bound among all
the UFB and CCB constraints. For future convenience, let us briefly
give the explicit form of this constraint.
By simple analytical minimization of the relevant terms of the
scalar potential it is possible to write the
value of all the {$\nu_{L_i},e_{L_j},e_{R_j}$} 
fields in
terms of the $H_u$ one. Then, for any value of $|H_u|<M_{GUT}$ satisfying
\bea
|H_u| > \sqrt{ \frac{\mu^2}{4\lambda_{e_j}^2}
+ \frac{4m_{L_i}^2}{g^{\prime 2}+g_2^2}}-\frac{|\mu|}{2\lambda_{e_j}} \ ,
\label{SU6}
\eea
the value of the potential along the UFB-3 direction is simply given
by
\bea
V_{\rm UFB-3}=(m_{H_u}^2
+ m_{L_i}^2 )|H_u|^2
+ \frac{|\mu|}{\lambda_{e_j}} ( m_{L_j}^2+m_{e_j}^2+m_{L_i}^2 ) |H_u|
-\frac{2m_{L_i}^4}{g^{\prime 2}+g_2^2} \ .
\label{SU8}
\eea
Otherwise
\bea
\label{SU9}
V_{\rm UFB-3}= m_{H_u}^2
|H_u|^2
+ \frac{|\mu|} {\lambda_{e_j}} ( m_{L_j}^2+m_{e_j}^2 ) |H_u| + \frac{1}{8}
(g^{\prime 2}+g_2^2)\left[ |H_u|^2+\frac{|\mu|}{\lambda_{e_j}}|H_u|\right]^2 \ .
\eea
In eqs.~(\ref{SU8}) and (\ref{SU9}) $\lambda_{e_j}$ is the leptonic Yukawa
coupling of the $j-$generation. 
Then, the
UFB-3 condition reads
\bea
\label{SU7}
V_{\rm UFB-3}(Q=\hat Q) > V_{\rm real \; min} \ ,
\eea
where $V_{\rm real \; min}=-\frac{1}{8}\left(g^{\prime 2} + g_2^2\right)
\left(v_u^2-v_d^2\right)^2$, with 
$v_{u,d}=\langle H_{u,d}^0\rangle$,
is the realistic minimum evaluated at the typical scale of
SUSY masses, say $M_{SUSY}$ 
(normally, as mentioned in the Introduction, a good
choice for $M_{SUSY}$ is an average of the stop masses),
and the renormalization scale $\hat Q$ is given by 
$\hat Q\sim {\rm Max}(\lambda_{top} |H_u|, M_{SUSY})$.
Notice from eqs.~(\ref{SU8}) and (\ref{SU9}) 
that the negative contribution to $V_{UFB-3}$
is essentially given by the $m_{H_u}^2$ term, which can be very sizeable in 
many instances. On the other hand, the positive contribution is dominated by 
the term $\propto 1/\lambda_{e_j}$, thus the larger
$\lambda_{e_j}$ the more restrictive
the constraint becomes. Consequently, the optimum choice of
the $e$--type slepton is the third generation one, i.e.
${e_j}=$ stau.


\section{
Experimental and astrophysical constraints}

As mentioned in the Introduction, 
we have to be sure that we obtain
a correct phenomenology in our analysis.
In this sense, we list here
the most recent experimental and astrophysical
results which are relevant when computing
the  neutralino-nucleon
cross section in the context of SUGRA.
They give rise to  
important constraints on the
parameter space.

\begin{itemize}

\item Higgs mass

Whereas in the context of the SM
the negative direct search for the Higgs at the LEP2 collider implies
a precise lower bound on its mass of 114.1 GeV,
the situation in SUSY scenarios is more involved.
In particular, in the framework of mSUGRA, 
one obtains \cite{Heinemeyer} for the lightest CP-even Higgs
$m_h \gsim 114.1$ GeV when $\tan\beta \lsim 50$,
and $m_h \gsim 91$ GeV when $\tan\beta$ is larger.
This is because 
the $ZZh$ coupling, which controls the detection
of the lightest MSSM Higgs at LEP, is 
$\sin^2(\alpha-\beta)\sim 1$ when $\tan\beta \lsim 50$, and 
a significant suppression of $\sin^2(\alpha-\beta)$
occurs only with
$\tan\beta > 50$. Recall in this sense that $\alpha$ is the Higgs mixing angle
in the neutral CP-even Higgs sector,
and $\sigma_{SUSY} (e^+ e^- \to Zh)=
\sin^2(\alpha-\beta)\sigma_{SM} (e^+ e^- \to Zh)$ \cite{sine}.
Let us remark anyway that generically $\tan\beta$ is constrained to be 
$\tan\beta\lsim 60$, since otherwise several problems arise. 
For example, for moderate values of $m$ and $M$,
the squared CP-odd Higgs mass, $m^2_A$, becomes negative,
unless a fine-tuning (in the sense that only certain
combinations of $m$ and $M$ are possible) is carried out.
In our computation below we will analyse values of $\tan\beta$ from
10 to 50.

Clearly, from the above discussion,
when the mSUGRA framework is relaxed 
$\sin^2(\alpha-\beta)$ must be computed for all points of the parameter
space in order to know which bound for 
the lightest MSSM Higgs must be applied \cite{barate}.



Let us finally remark that
we evaluate $m_h$ using the 
program {\tt FeynHiggsFast},
a simplified version of the updated program {\tt FeynHiggs} \cite{FeynHiggs}
which contains the complete one-loop and dominant two-loop corrections.
The value of $m_h$ obtained with {\tt FeynHiggsFast} is
approximately 1 GeV below the one obtained using {\tt FeynHiggs}.
In addition, although the value of $m_h$ obtained with FeynHiggs
has an uncertainty of about 3 GeV, due e.g. to higher-order
corrections,
we have no taken it into account in our computation.

\item Top mass

Needless to say we use as input for the
top mass throughout this paper 
the central experimental value $m_t(pole)=175$ GeV.
However, let us remark that a modification in this mass by
$\pm 1$ GeV implies, basically, a modification also of $\pm 1$ GeV
in the value of $m_h$ obtained here.

\item Bottom and tau masses


For the bottom mass we take as input 
$m_b(m_b)=4.25$ GeV, which,  
following the analysis of ref.~\cite{santamaria}
with $\alpha_s(M_Z)=0.1185$,
corresponds to $m_b(M_Z)=2.888$
GeV.
In the evolution of the bottom mass we incorporate the SUSY
threshold corrections \cite{pierce} at $M_{SUSY}$.
These are known to be significant, specially for large
values of $\tan\beta$. 
We also follow a similar analysis for the 
tau mass, using as input
$m_{\tau}(M_Z)=1.7463$ GeV.

\item SUSY spectrum

We impose
the present experimental lower
bounds on SUSY masses coming from LEP and Tevatron. In particular,
using the low-energy relation from mSUGRA,
$M_1=\frac{5}{3}\tan^2\theta_W M_2$,
one obtains for the lightest chargino mass the bound \cite{chargino}
$m_{\tilde\chi_1^{\pm}}>103$ GeV.
Likewise, one is also able to obtain the following 
bounds for sleptons masses \cite{sleptons}:
$m_{\tilde e}>99$ GeV,
$m_{\tilde\mu}>96$ GeV,
$m_{\tilde\tau}>87$ GeV.
Finally, we use the following 
bounds on the masses of the 
sneutrino, the lightest stop, the rest of squarks, and
gluinos: 
$m_{\tilde\nu}>50$ GeV,
$m_{\tilde t}>95$ GeV,
$m_{\tilde q}>150$ GeV,
$m_{\tilde g}>190$ GeV.




\item $b\to s\gamma$

The measurements of $B\to X_s\gamma$ decays 
at 
CLEO \cite{cleo} 
and BELLE \cite{belle},
lead to bounds on the branching ratio 
$b\to s\gamma$. In particular we impose on our computation
$2\times 10^{-4}\leq BR(b\to s\gamma)\leq 4.1\times
10^{-4}$, where the evaluation 
is carried out using the routine provided by
the program {\tt micrOMEGAs} \cite{micromegas}. 
A description of this procedure can be found in 
ref.~\cite{routine}. Although the improvements
of ref.~\cite{impro} are not included in this routine,
they are not so important for our study since we consider
only $\mu>0$.




\item $g_{\mu}-2$

The new measurement of the anomalous
magnetic moment of the muon, $a_\mu=(g_{\mu}-2)/2$, 
in the E821 experiment at the 
BNL \cite{muon} 
deviates by 
$(33.7\pm 11.2) \times 10^{-10}$
from the recent SM calculation of ref.~\cite{news} using
$e^+e^-$ data.
Assuming that the possible 
new physics is due to SUGRA, we will show in our computation the 
constraint
$11.3\times 10^{-10}\leq a_{\mu} (SUGRA)\leq 56.1\times
10^{-10}$ at the
2$\sigma$ level. This would exclude the case $\mu<0$.
However, it is worth noticing that the above result is in contradiction 
with the one obtained by using 
tau decay data (instead of $e^+e^-$ ones) which only imply a deviation
$(9.4\pm 10.5) \times 10^{-10}$ from the SM calculation \cite{news}.


\item LSP


The lightest supersymmetric particle (LSP) must be an electrically neutral
(also with no strong interactions) particle,
since otherwise it 
would bind to nuclei and would be excluded as a candidate
for dark matter from unsuccessful searches for exotic heavy 
isotopes \cite{isotopes}.
Although the $\tilde{\chi}_1^0$ is
the LSP in most of the parameter space of SUGRA,
in some regions
one of the staus, $\tilde{\tau}_1$, can be lighter.
Therefore, following the above argument,
we discard these regions.

\item Relic $\tilde{\chi}_1^0$ density

We impose in general
the preferred astrophysical bounds on the dark matter density,
$0.1\lsim \Omega_{DM}h^2\lsim 0.3$,
on our 
theoretical computation of the relic $\tilde{\chi}_1^0$ density,
assuming $\Omega_{DM}\approx \Omega_{\tilde{\chi}_1^0}$.
For the sake of completeness we also show in the figures the bounds
$0.094\lsim \Omega_{DM}h^2\lsim 0.129$ deduced from the recent data
obtained by the WMAP satellite \cite{wmap}.
Let us remark that the theoretical computation of the
relic density 
depends on assumptions about the evolution of the early
Universe, and therefore different cosmological scenarios give
rise to different results \cite{relic}.
We will consider in our analysis the standard mechanism of 
thermal production of neutralinos.

Let us finally mention that we evaluate $\Omega_{\tilde{\chi}_1^0}$
using the program {\tt microMEGAs} \cite{micromegas}.
The exact tree-level cross sections for all possible annihilation \cite{kami}
and coannihilation \cite{stau,char,stop} 
channels are included in the code through a link
to {\tt CompHEP} \cite{comphep}, and accurate thermal average of them is used.
Also, poles and thresholds are properly handled and one-loop
QCD corrected Higgs decay widths \cite{width}
are used. The SUSY corrections included in the latest version of the
code \cite{width} are not implemented yet by {\tt micrOMEGAs}. Fortunately,
in our case, their effect is much smaller than that of the QCD
corrections.
Good agreement between {\tt micrOMEGAs} and other independent
computations of $\Omega_{\tilde{\chi}_1^0}$ including
$\tilde{\chi}^0_1-\tilde{\tau}_1$ coannihilations can be found in 
ref.~\cite{compare}.



\end{itemize}

\section{mSUGRA predictions for the neutralino-nucleon cross section}

Let us recall first that the relevant effective Lagrangian describing
the elastic scattering of  $\tilde{\chi}^0_1$ on protons and neutrons
has a spin-independent (scalar) interaction and a spin-dependent
interaction \cite{kami}.
However, the contribution of the scalar one is generically larger and
therefore we will concentrate on it. 
This scalar interaction includes contributions from squark exchange and
neutral Higgs exchange.
Let us also remark that
the scalar cross sections for both, protons and neutrons, 
are basically equal.
Taking all the above into account, in this section 
we will compute the predictions for the
scalar neutralino-proton cross section
$\sigma_{\tilde{\chi}_1^0-p}$ in the context of mSUGRA,
where the soft terms are assumed to be universal. 
As discussed in the Introduction, this is the simplest possibility
and may arise in specific string constructions.
In particular we will carry out the analysis  
following first the usual assumption of a GUT scale,
$M_{GUT} \approx 2\times 10^{16}$ GeV, 
and later we will modify this assumption allowing the
possibility of an intermediate scale.

\subsection{GUT scale}

\begin{figure}
\hspace*{-1.2cm}\epsfig{file=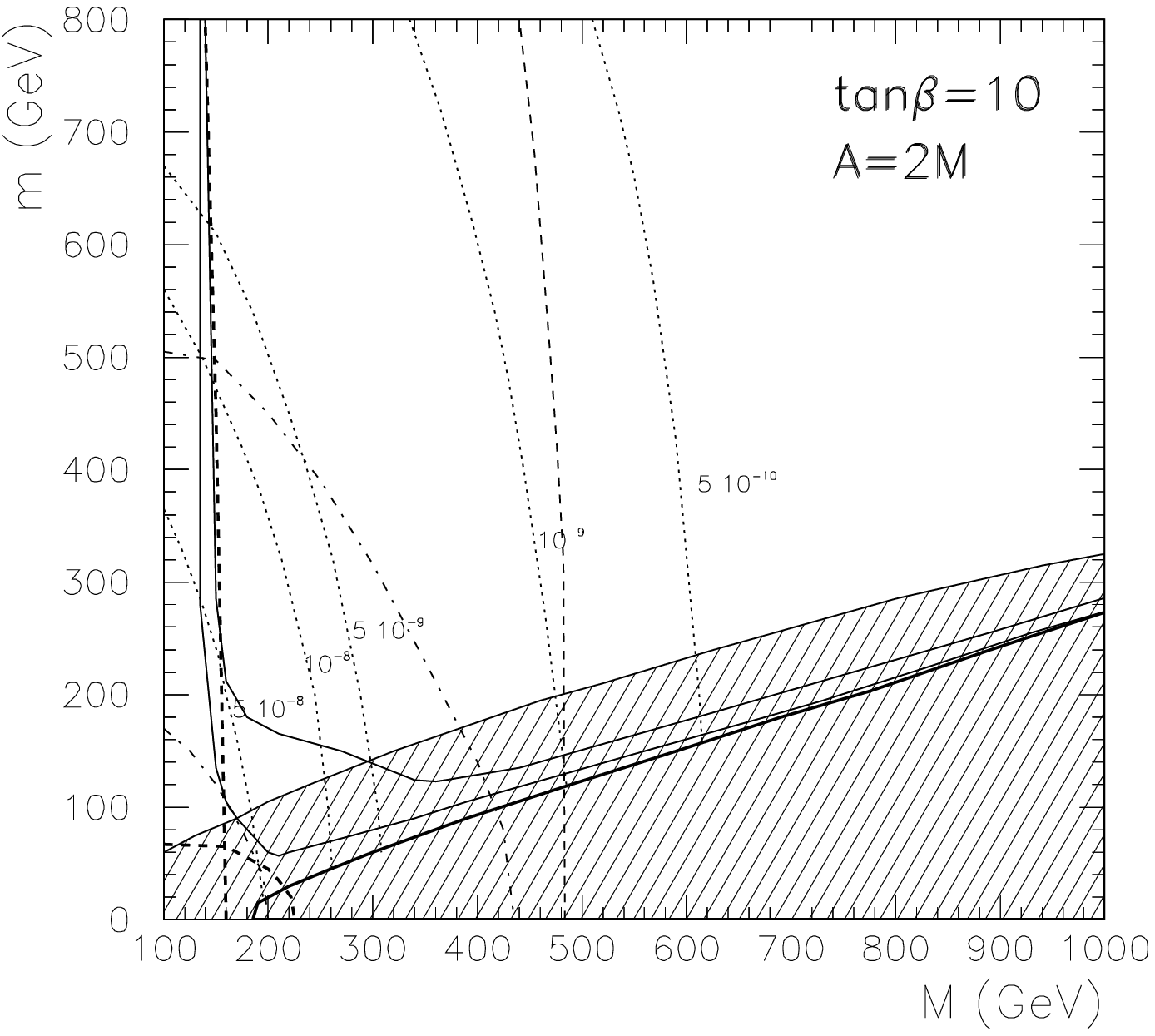,width=9cm}
%
\epsfig{file=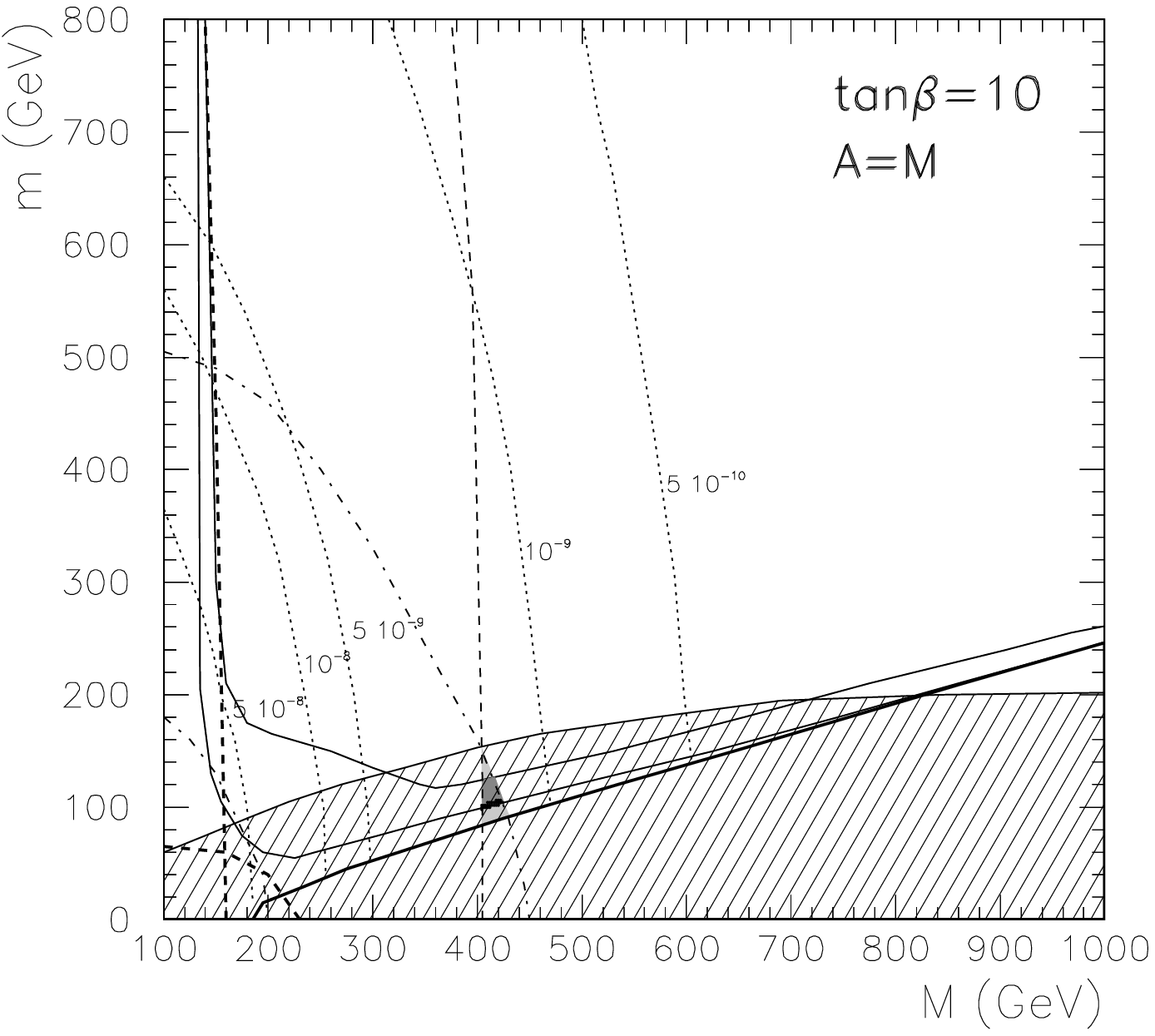,width=9cm} 

\hspace*{-1.2cm}\epsfig{file=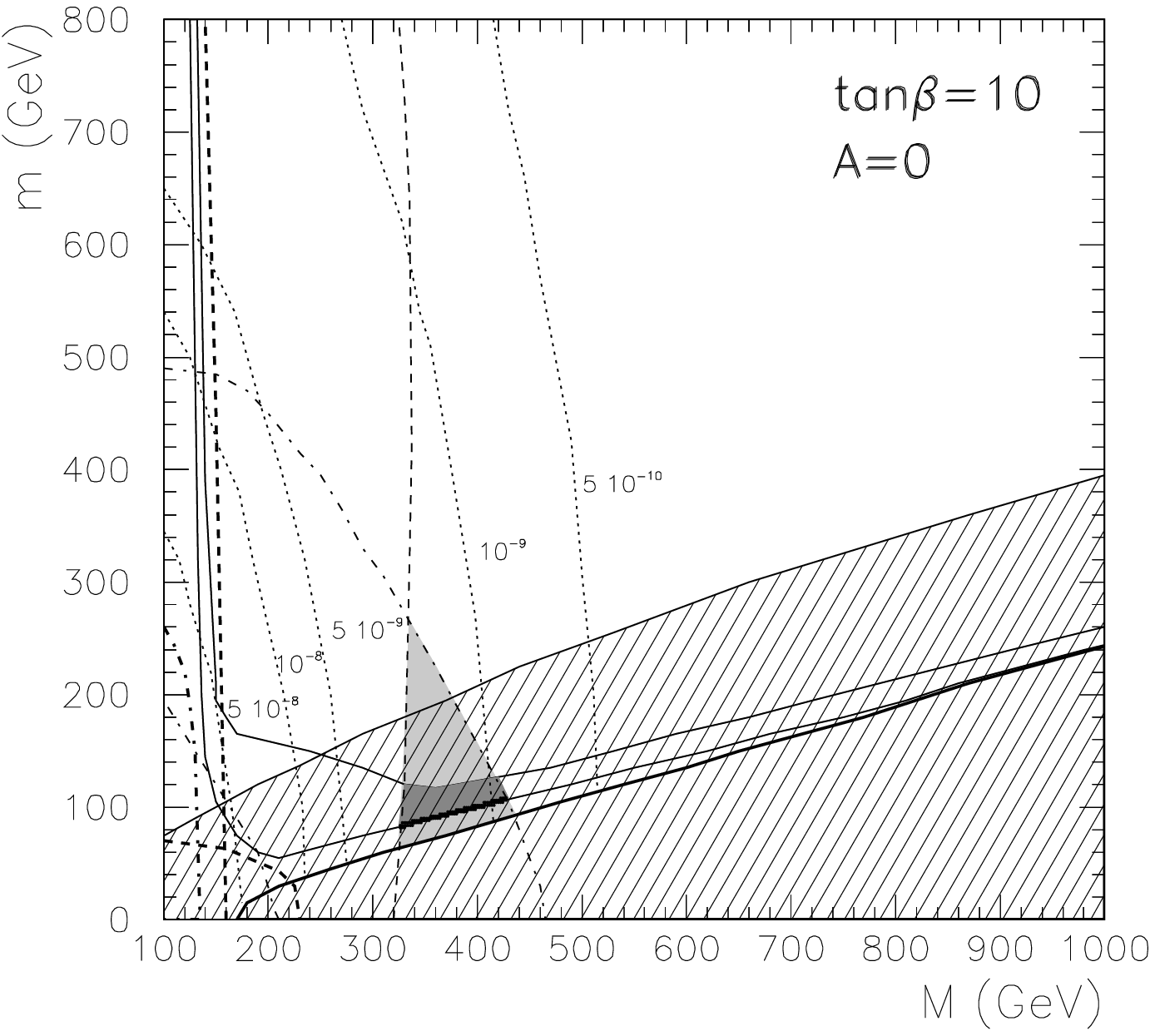,width=9cm}
\epsfig{file=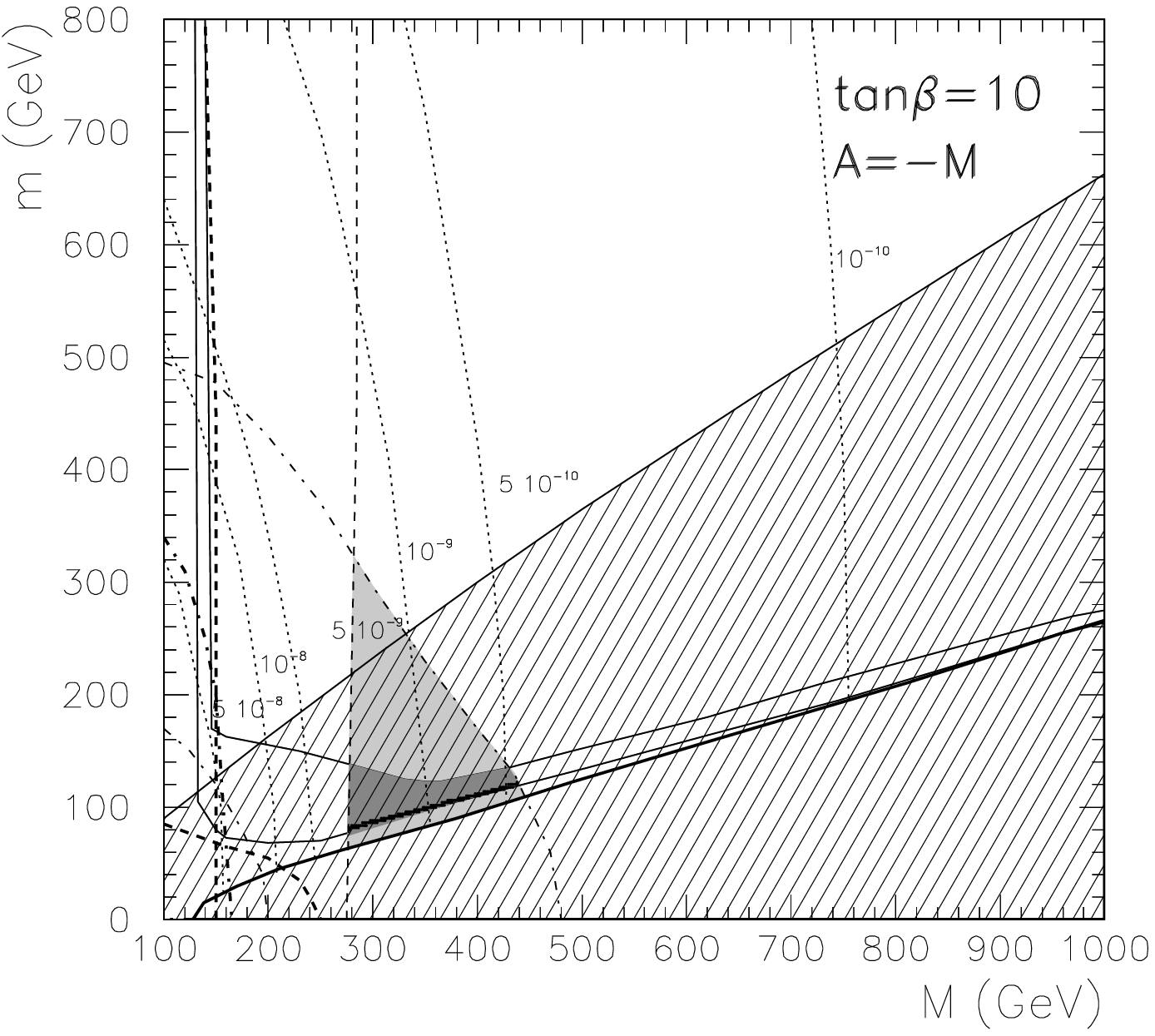,width=9cm}


\captions{Scalar neutralino-proton cross section $\sigma_{\tilde{\chi}_1^0-p}$
in the parameter space of the mSUGRA scenario ($m$, $M$) 
for $\tan \beta=10$
and $\mu > 0$, using different values of $A$.
The dotted curves are contours of $\sigma_{\tilde{\chi}_1^0-p}$.
The region to the left of the
near-vertical dashed line is excluded by the lower bound 
on the Higgs mass $m_h>114.1$ GeV. 
The region to the left of the near-vertical double dashed line
is excluded by the lower bound on the chargino mass
$m_{\tilde\chi_1^{\pm}}>103$ GeV.
The corner in the lower left shown also by a double dashed line
is excluded by the LEP bound on the stau mass
$m_{\tilde{\tau}_1}>87$ GeV.
The region bounded by dot-dashed lines is allowed by $g_{\mu}-2$.
The region to the left of the 
double dot-dashed line is excluded by $b\to s\gamma$.
From bottom to top, the solid lines are the upper bounds of the areas such
as $m_{\tilde{\tau}_1}<m_{\tilde{\chi}_1^0}$ (double solid), 
$\Omega_{\tilde{\chi}_1^0} h^2<0.1$ and $\Omega_{\tilde{\chi}_1^0} h^2<0.3$. 
The light shaded area is favored by all the phenomenological
constraints, 
while the dark one fulfils in addition
$0.1\leq \Omega_{\tilde{\chi}_1^0}h^2\leq 0.3$ (the black region on top of this
indicates the WMAP range $0.094<\Omega_{\tilde{\chi}_1^0}h^2<0.129$).
The ruled region 
is excluded because of
the charge and colour breaking constraint UFB-3.
\label{a2m}}

\end{figure}

As is well known, 
in the mSUGRA scenario with a GUT scale
$\tilde{\chi}^0_1$ is mainly bino
and, as a consequence, the predicted $\sigma_{\tilde{\chi}_1^0-p}$
is well below the accessible experimental regions for low and moderate
values of $\tan\beta$.
We show this fact in
Fig.~\ref{a2m}, where contours 
of $\sigma_{\tilde{\chi}_1^0-p}$ in the
parameter space ($m$, $M$) for $\tan \beta=10$
and $\mu > 0$ are plotted for different values of $A$.
We choose $A$ proportional to $M$ because this relation is 
particularly interesting, arising naturally in several string 
models \cite{dilaton}. However our conclusions will be independent on 
this assumption. 
For example, if we choose to do the plots for different constant 
values of $A$, a very common procedure in pure SUGRA analyses,
the results will be qualitatively similar.
Let us also remark that
the sign of the dominant contribution to the supersymmetric       
contribution to the muon anomalous magnetic moment is given by $M_2\mu$.
As we discussed in the previous section, we 
are taking this to be positive,
and therefore we will only consider $sign(M)=sign(\mu)$. Now, due to the
symmetry of the RGEs, the results for ($-M,A,-\mu$) are identical to those
for ($M,-A,\mu$), and therefore, we can cover the whole (permitted)
parameter space restricting to positive values for $M$ and $\mu$ and
allowing $A$ to take positive and negative values.
On the other hand,
note that we can deduce the value of the $\tilde{\chi}_1^0$ mass in the
plots from the value of $M$, since $\neumass\sim 0.4\ M$. 
For the gluino mass we can also use the simple relation,
$m_{\tilde g}\sim 2.5\ M$.



As we can see in the figure, the experimental bounds discussed in
Section~3 are very important and exclude large regions of the
parameter
space. For example, for $A=2M$ the whole parameter space is forbidden.
This is due to the combination of the Higgs mass bound with
the $g_{\mu}-2$ lower bound (recall that we are using only the limit
based on $e^+e^-$ analysis). 
Notice in this sense that
the bound on the Higgs mass is 
less restrictive for smaller  
values of $A$, and therefore the allowed region is relatively
large for $A=0,-M$.
For example, for $A=0$ one obtains from the Higgs mass 
the lower bound $M\gsim$ 320 GeV, and from $g_{\mu}-2$ the upper bound
$M\lsim$ 440 GeV. These bounds imply for the neutralino,
$128\lsim m_{\tilde{\chi}_1^0}\lsim 176$ GeV, and for the
gluino (and squarks)
$800\lsim m_{\tilde{g}}\lsim 1100$ GeV. 
The light shaded area in Fig.~1 shows the
region allowed by the experimental bounds. There, the lower contour
is obtained including also
the constraint coming from the LSP bound.
On the other hand, when the astrophysical bounds
$0.1\lsim \Omega_{\tilde{\chi}_1^0}h^2\lsim 0.3$
are
also imposed the allowed area becomes very small 
(extremely small if the recent WMAP data are taken into account). Only
the beginning of the tail where the LSP is almost degenerate with
the stau, producing efficient coannihilations, is rescued.
For $A=-M$ a small part of the bulk region with moderate
$m$ and $M$ is also rescued.
For this area $\sigma_{\tilde{\chi}_1^0-p}\approx 10^{-9}$ pb.

In addition, 
the restrictions coming from
the UFB-3 constraint discussed in Section~2
exclude also this area.
We have checked that this is also true for smaller values of $A$,
as e.g. $A=-2M$. 
Let us remark that this constraint depends crucially on the value
of $A$. In particular the UFB-3 excluded (ruled) region is larger
for negative values of $A$ and also increases with $|A|$ as shown
in Fig.~1.
This can be understood from the evolution of $m_{H_u}^2$ with the
scale, since for the above cases it becomes more negative.
From the discussion concerning eqs.~(\ref{SU8}) and (\ref{SU9})
we deduce that the negative contribution to $V_{UFB-3}$
is more important and therefore the UFB-3 minima are easily
deeper than the realistic one. Note that the region excluded 
because the stau is the LSP is also generically excluded by
the UFB-3 constraint. 
Again, this can be understood 
from eqs.~(\ref{SU8}) and (\ref{SU9}), since the positive contribution
to $V_{UFB-3}$ depends on the stau mass, and this is now small.
In conclusion, the results indicate that the whole parameter
space for $\tan\beta=10$
is excluded
on these grounds.
It is worth noticing, however, that when 
the $g_{\mu}-2$ lower bound is relaxed, for large
values of $M$ a narrow region of the
area fulfilling the astrophysical bounds
might be allowed for the case $A=M$.

\begin{figure}

\hspace*{-1.2cm}\epsfig{file=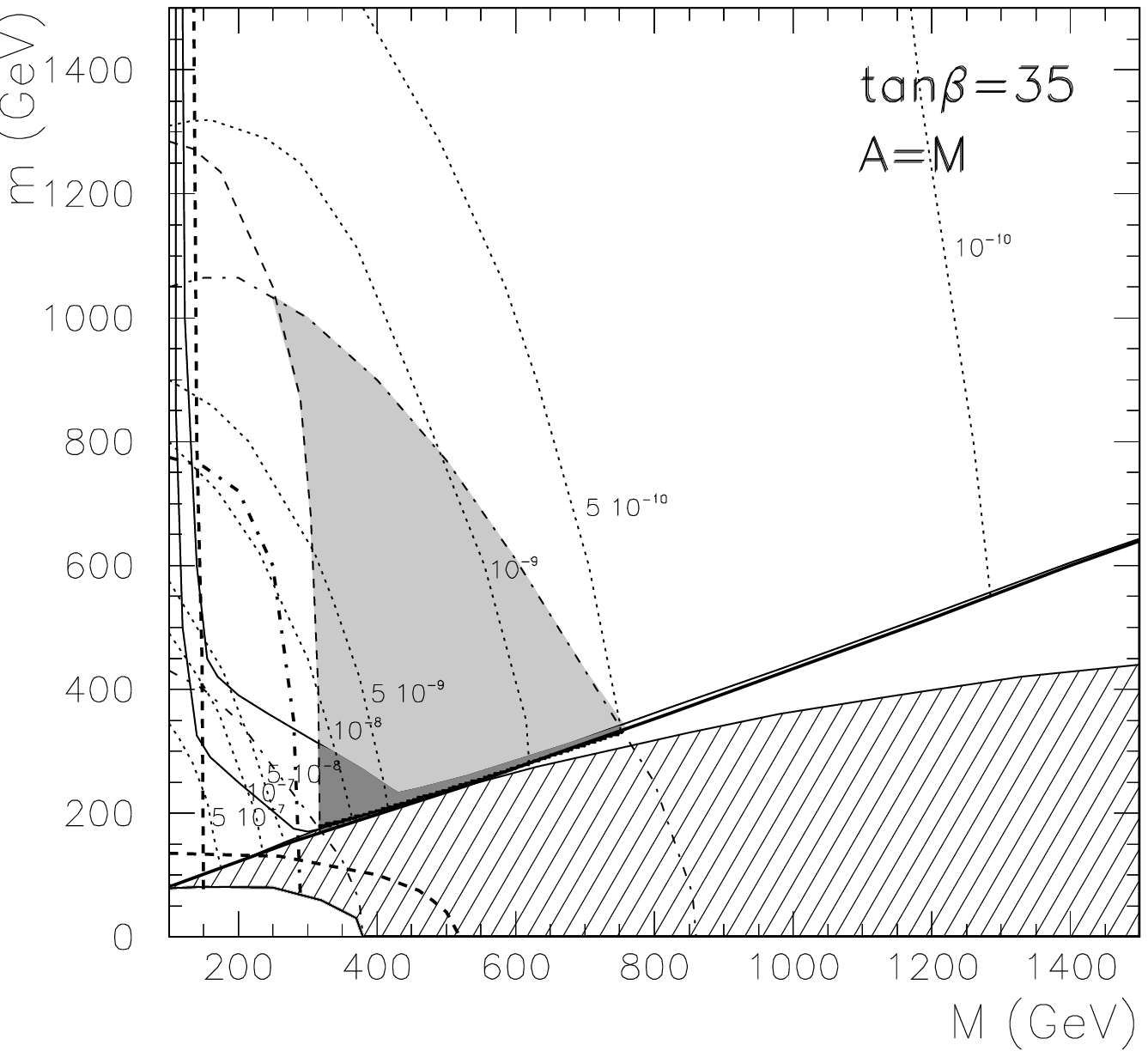,width=9cm}
%
\epsfig{file=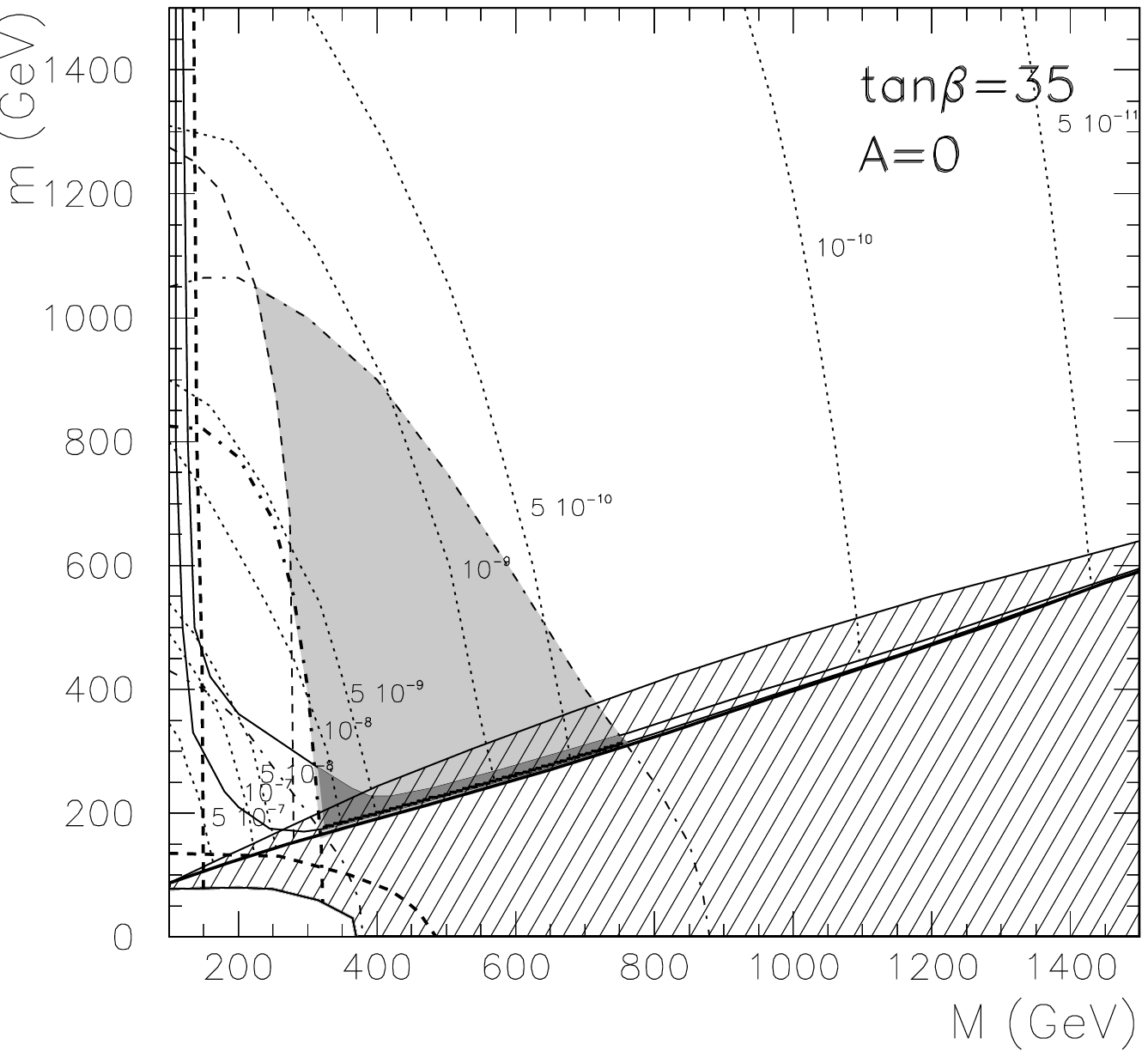,width=9cm}

\hspace*{-1.2cm}\epsfig{file=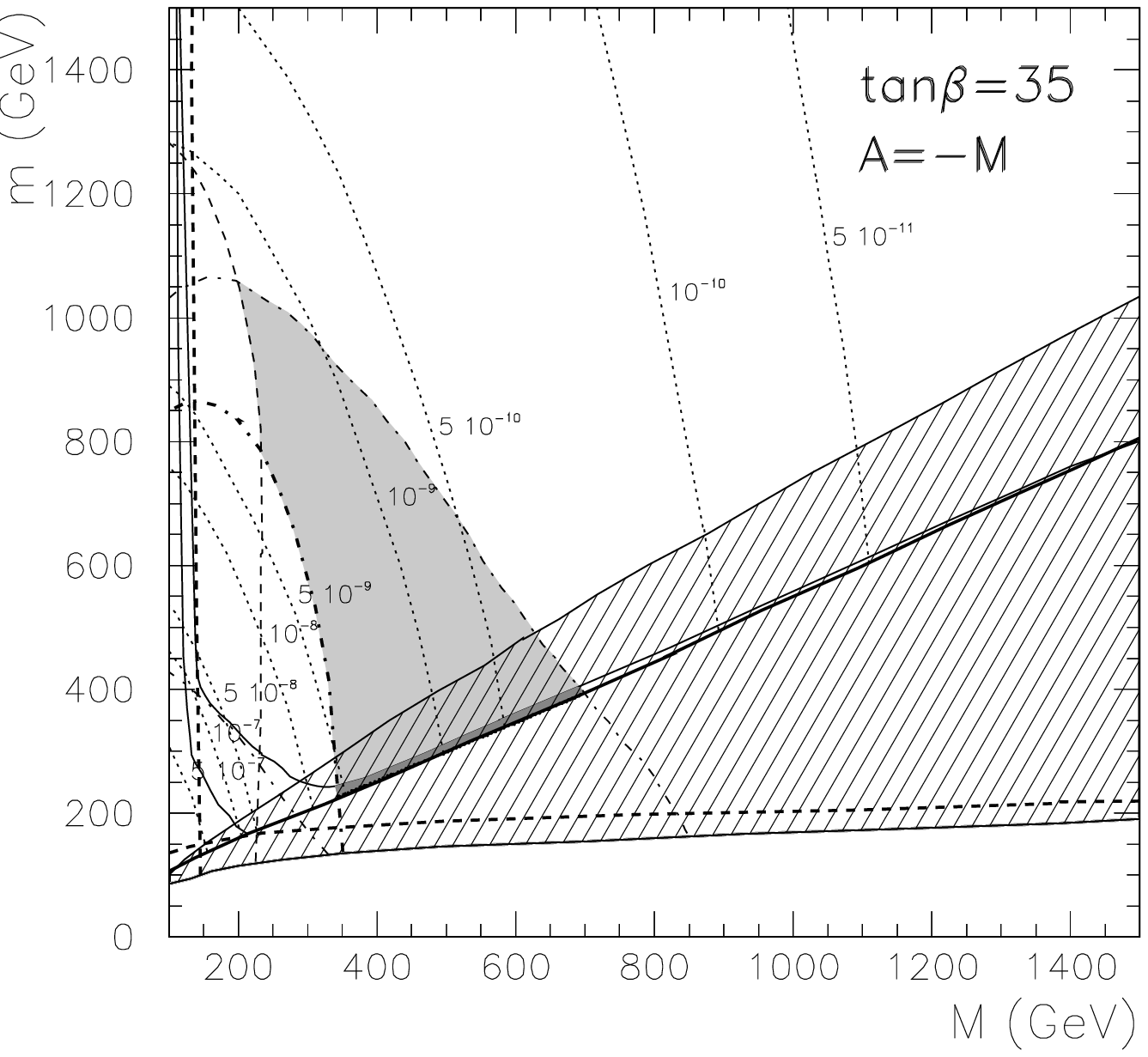,width=9cm}
%
\epsfig{file=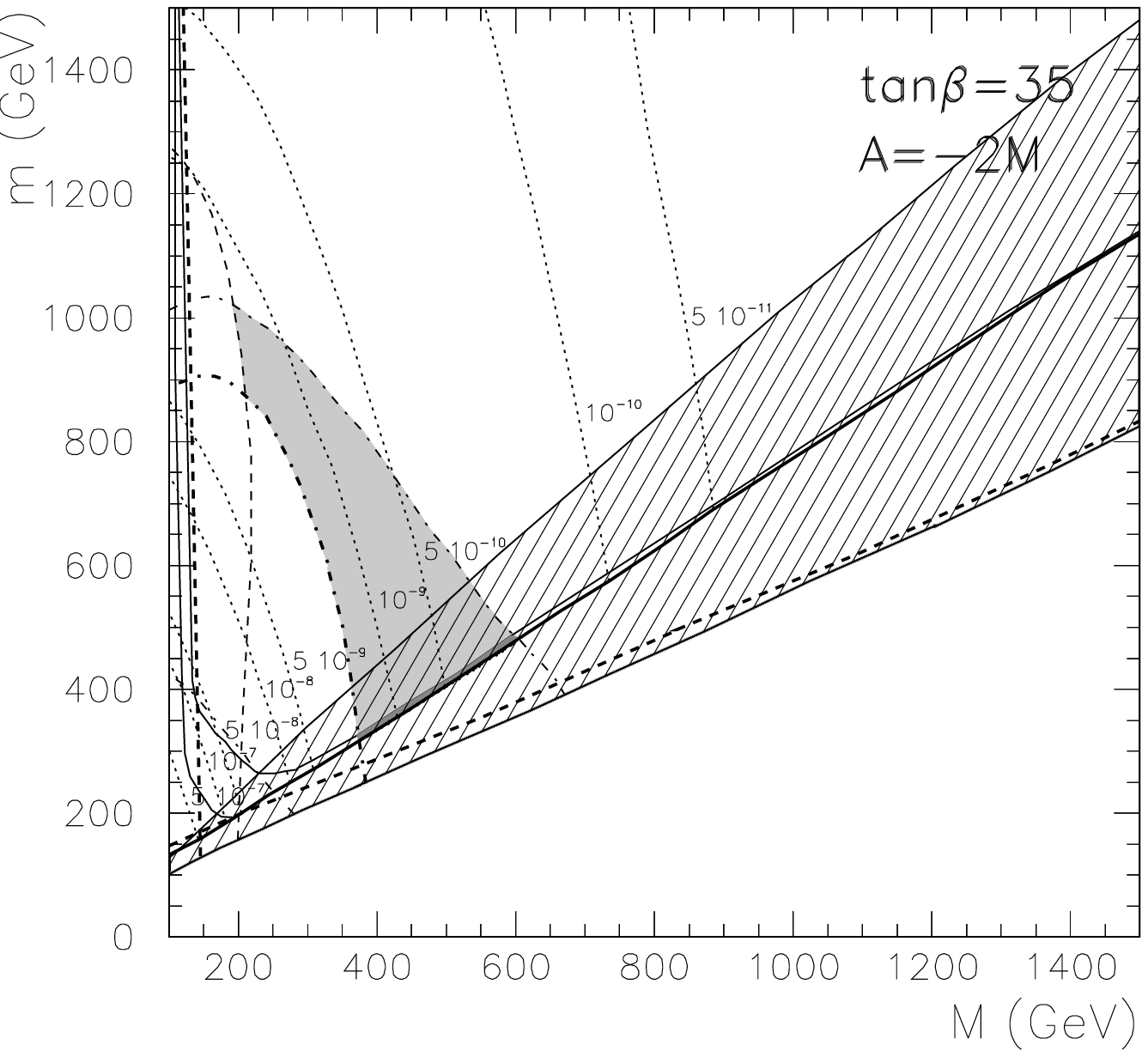,width=9cm}

\captions{The same as in Fig.~\ref{a2m} but for $\tgb=35$ and 
different values of $A$.
The white region at the bottom bounded by a solid line
is excluded because 
$m_{\tilde{\tau}_1}^2$ becomes negative.
\label{a35}}
\end{figure}

Although
the cross section increases, entering in the DAMA 
region \cite{experimento1,halo}, when the value
of $\tan\beta$ increases \cite{Bottino,Arnowitt}, 
the present experimental constraints 
exclude this possibility \cite{Ellis,Arnowitt3}.
We show this fact in
Fig.~\ref{a35} for $\tan \beta=35$ and four representative values of
$A$. 
In principle, 
if we only impose the LEP lower bound $m_{\tilde\chi_1^{\pm}}>103$ GeV,
the cross section can be as large as 
$\sigma_{\tilde{\chi}_1^0-p}\approx 10^{-6}$ pb.
However, at the end of the day,
the 
other 
experimental bounds (Higgs mass, $b\to s\gamma$,
$g_{\mu}-2$ upper bound) 
constrain the cross section  
to be $\sigma_{\tilde{\chi}_1^0-p}\lsim 10^{-8}$ pb.
The region allowed by 
the $g_{\mu}-2$ lower bound is now larger and the cross section 
can be as low as
$\sigma_{\tilde{\chi}_1^0-p}\gsim 10^{-10}$ pb.
For example, for $A=0$ (and similarly for the others) these bounds imply 
$260\lsim M\lsim 750$ GeV, and therefore 
$104\lsim m_{\tilde{\chi}_1^0}\lsim 300$ GeV, 
$650\lsim m_{\tilde{g},\tilde{q}}\lsim 1875$ GeV.

Concerning the UFB-3 constraint,
it is worth noticing, by comparison with Fig.~1, that
the larger $\tgb$ is,
the larger 
the excluded region becomes. 
This is because
the positive contribution for $\ufbiii$ 
is multiplied 
by the inverse of the tau Yukawa coupling.
Since this coupling is proportional to $\frac1{\cos\beta}$,
it increases when $\tgb$ increases
leading to a decrease in the positive 
contribution. As a consequence
the UFB-3 constraint becomes more restrictive.
However, unlike the case 
$\tan \beta=10$,
this is not sufficient to forbid
the whole dark shaded area allowed also by
the astrophysical bounds,
because the increase in the tau coupling
also produces an increase of the region where
the stau is the LSP, and therefore the coannihilation tail is also risen.
This is very clear for $A=M$, where the region forbidden
by the LSP bound is even larger than the one forbidden by
the UFB-3.
Since the area bounded by solid lines, fulfilling 
$0.1\leq \Omega_{\tilde{\chi}_1^0}h^2\leq 0.3$,
has the lower edge above and very close to the LSP 
line (recall that coannihilations are very important in this region
producing the correct amount of relic neutralino density),
we can always find values of the parameters where 
all constraints are fulfilled.
In fact, one can check that 
$\tan\beta > 20$ is needed to obtain this result.
In any case, it is worth noticing that, even for these large values of 
$\tan\beta$, the UFB-3 constraint can be very important
depending on the value of $A$. For example, in Fig.~2 a large region
of the dark shaded area turns out to be forbidden when $A=0$,
and the whole region when $A=-M,-2M$.

\begin{figure}

\hspace*{-1.2cm}\epsfig{file=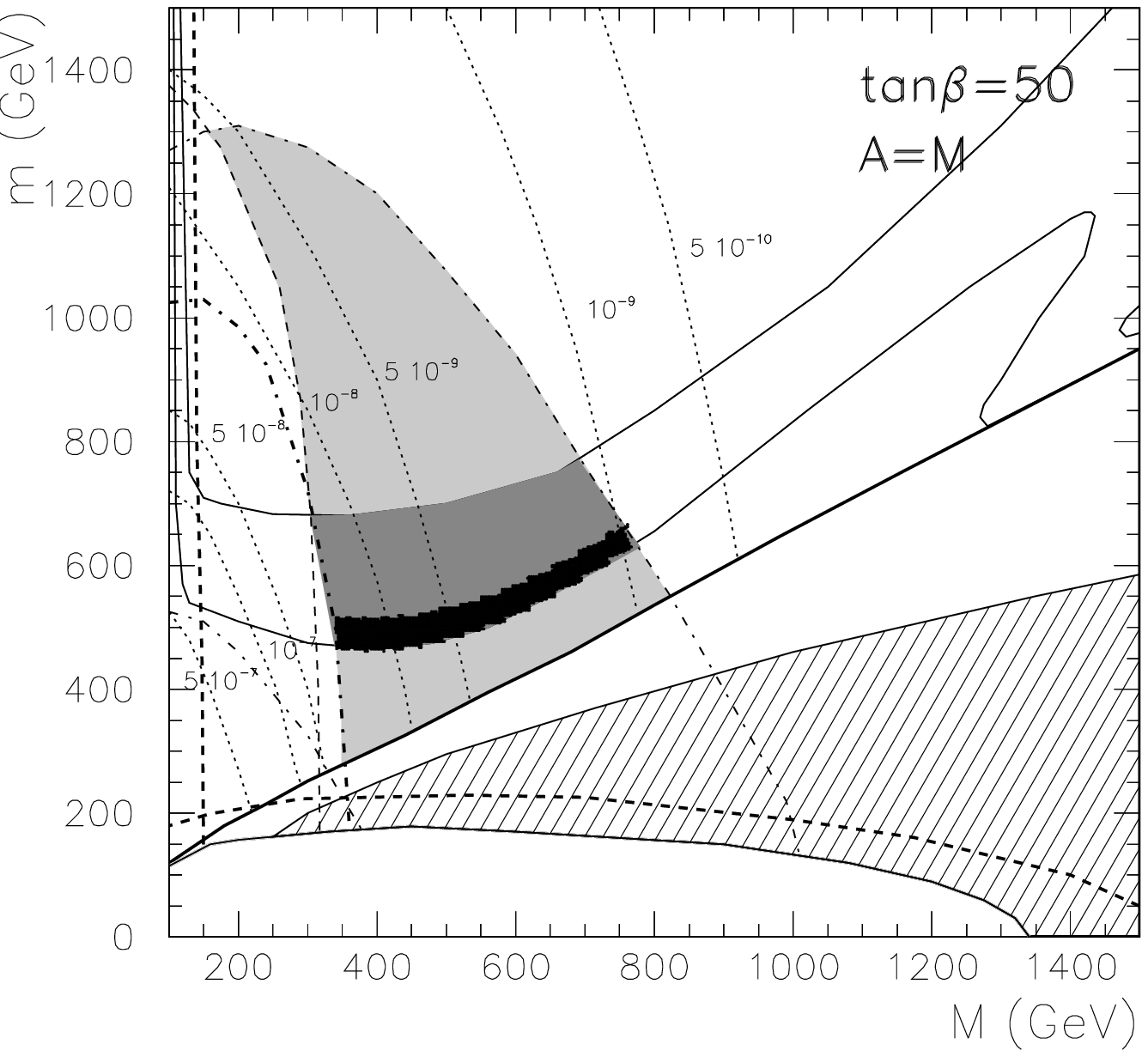,width=9cm}
%
\epsfig{file=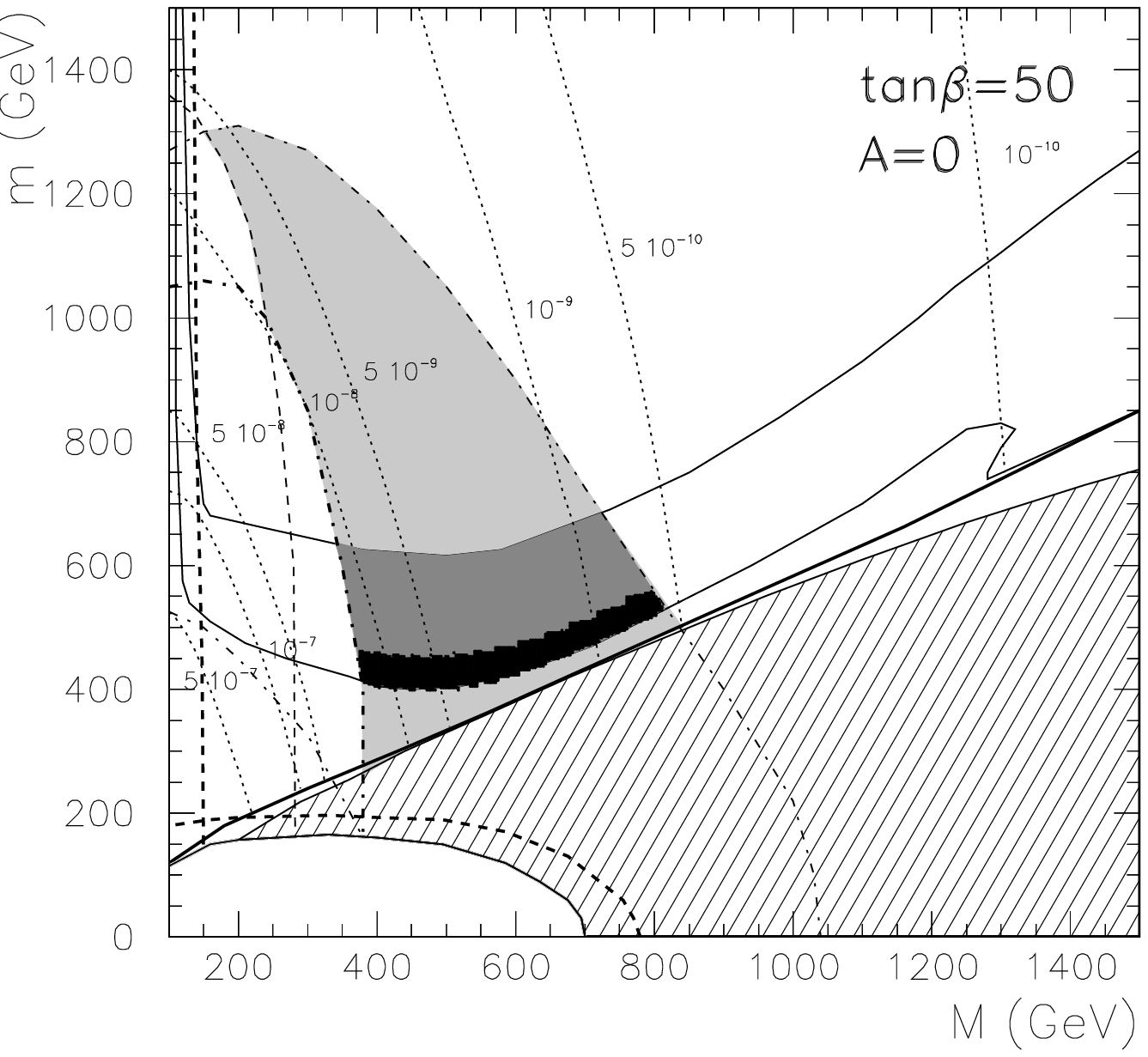,width=9cm}

\hspace*{-1.2cm}\epsfig{file=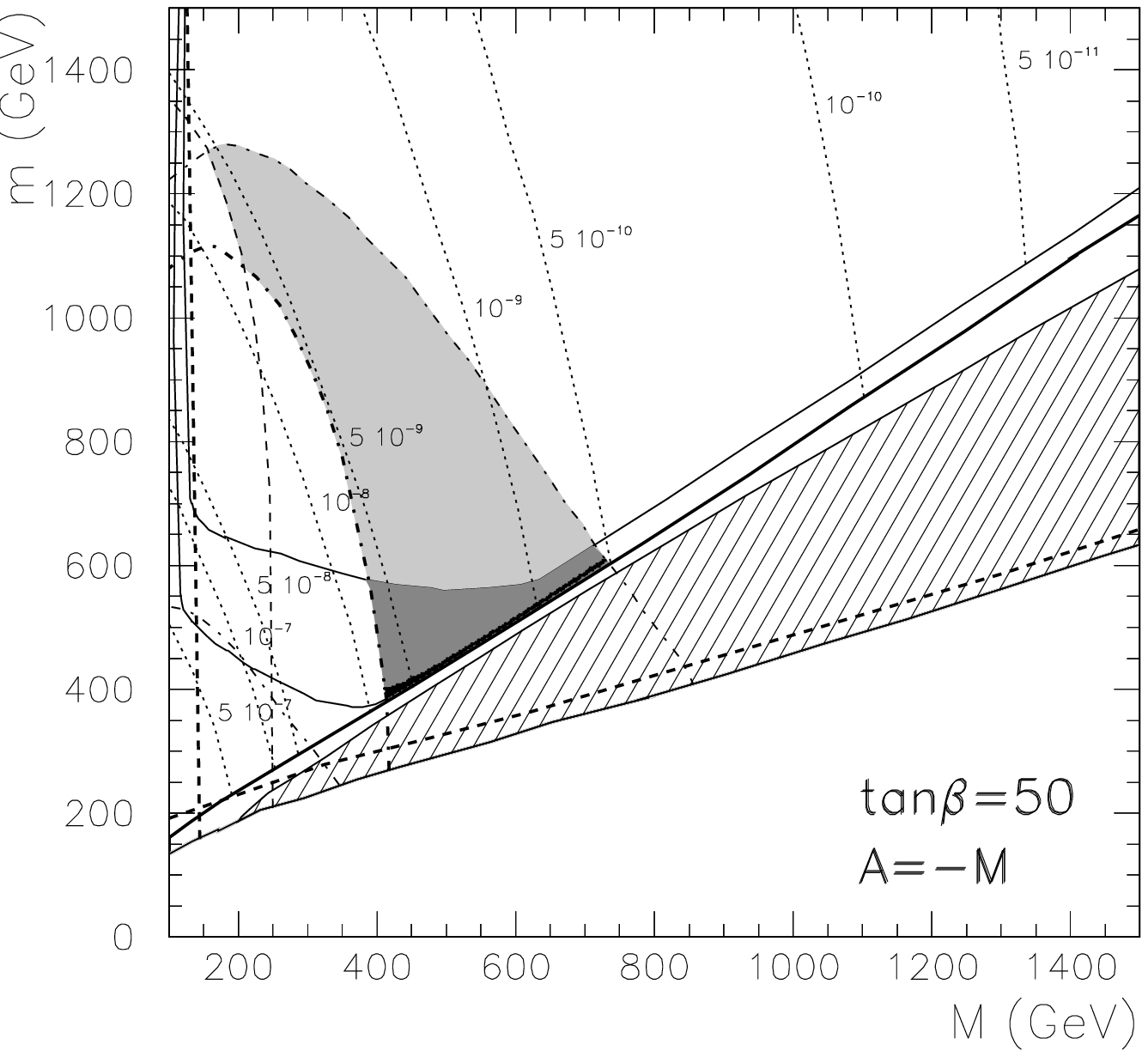,width=9cm}
%
\epsfig{file=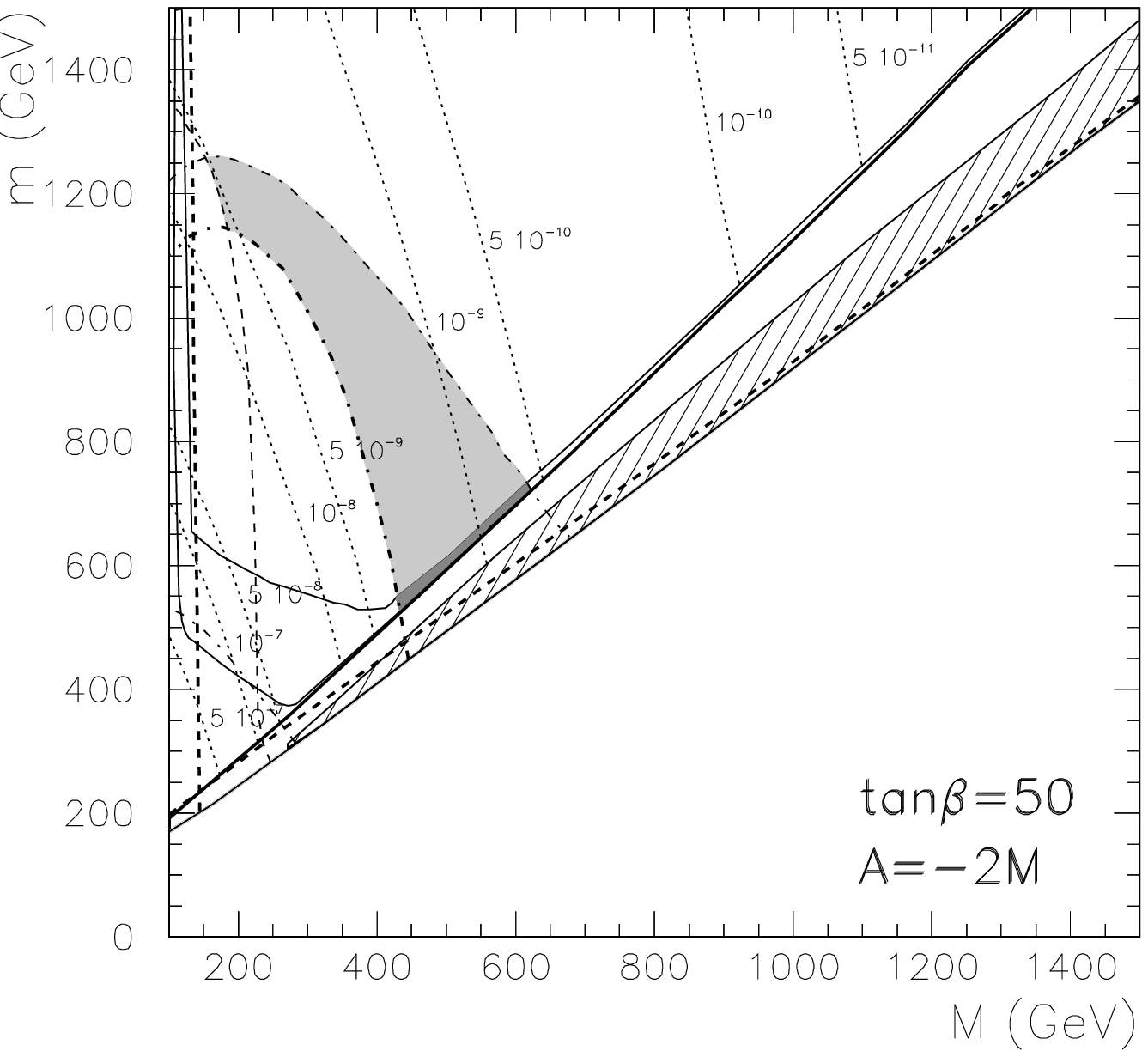,width=9cm}

\captions{The same as in Fig.~\ref{a2m} but for $\tgb=50$ and
different values of $A=M$.
The white region at the bottom bounded by a solid line
is excluded because 
$m_{\tilde{\tau}_1}^2$ becomes negative.
\label{a355}}
\end{figure}

The above comments can also be applied for very large values of 
$\tan\beta$, as e.g. $\tan \beta=50$.
We show this in Fig.~\ref{a355}. Note that now
for $A=0,-M,-2M$, unlike the case $\tan \beta=35$, 
the whole dark shaded areas allowed by experimental
and astrophysical bounds are not constrained by the UFB-3.
Let us finally recall that the region allowed by 
$0.1\leq \Omega_{\tilde{\chi}_1^0}h^2\leq 0.3$
is larger because the 
CP-odd Higgs $A$ becomes lighter as $\tan\beta$ increases.
This allows the presence of resonances in the Higgs mediated
annihilation channels, resulting in drastic reduction
of the neutralino relic abundance.
In the case of $\tan\beta=50$, the resonant effects in the
annihilation channels are felt in the whole parameter
space displayed in Fig.~2.
We can see as well, that for $A=M,0$ the area of the parameter space where
$\tilde{\chi}_1^0-\tilde{\tau}_1$ 
coannihilations are relevant lead 
to values of 
$\Omega_{\tilde\chi^0_1} h^2<0.1$.

\begin{figure}
\hspace*{-1.2cm}\epsfig{file=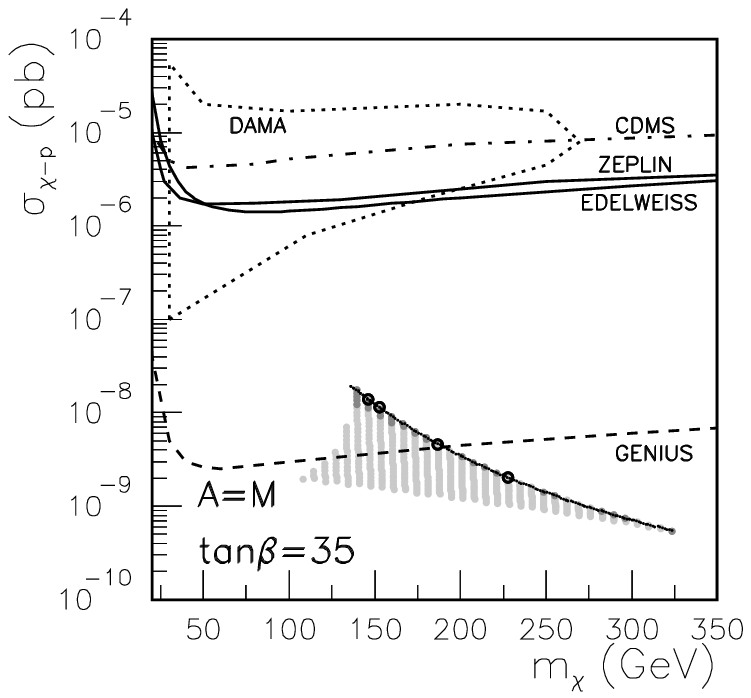,width=9cm}
%
\epsfig{file=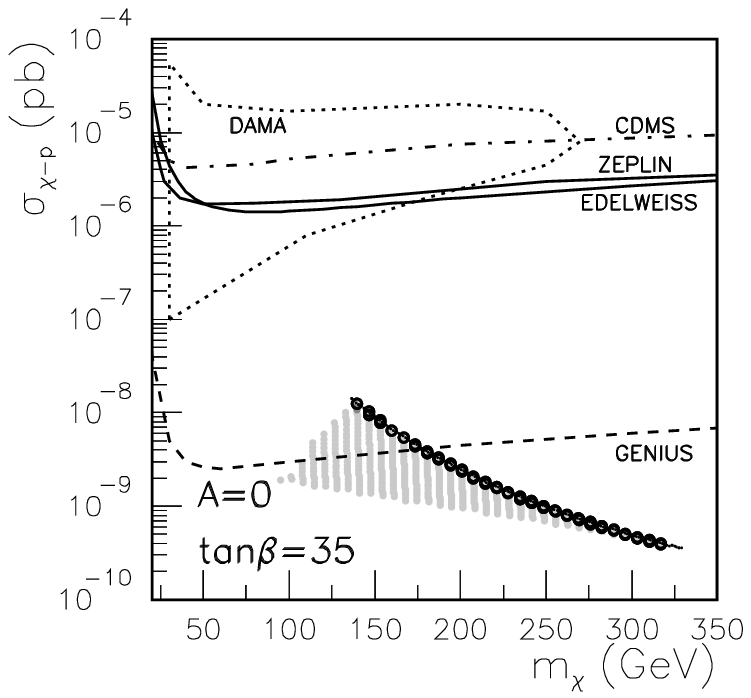,width=9cm}
%
\captions{Scatter plot of 
the scalar neutralino-proton cross section
$\sigma_{\tilde{\chi}_1^0-p}$ 
as a function of 
the neutralino mass 
$m_{\tilde{\chi}_1^0}$ in the mSUGRA scenario,
for
$\tan\beta=35$ and $A=M,0$,
with $\mu>0$. The light grey dots correspond to points
fulfilling all experimental constraints.
The dark grey dots correspond to points fulfilling in addition
$0.1\leq \Omega_{\tilde{\chi}_1^0}h^2\leq 0.3$
(the black dots on top of these
indicate those fulfilling the WMAP range $0.094<\Omega_{\tilde{\chi}_1^0}h^2<0.129$). 
The circles indicate regions 
excluded by the UFB-3 constraint.
The area bounded by dotted lines is allowed by the DAMA 
experiment \cite{experimento1,halo}. The (upper) areas
bounded by dot-dashed and solid lines are
excluded by CDMS \cite{experimento2} and EDELWEISS \cite{edelweiss} 
and ZEPLIN I \cite{zeplin1} current experimental
limits.
The (upper) area bounded by the dashed line will be analized by the 
projected GENIUS experiment \cite{GENIUS}.
\label{cross_scale122}}
\end{figure}

\begin{figure}
\hspace*{-1.2cm}\epsfig{file=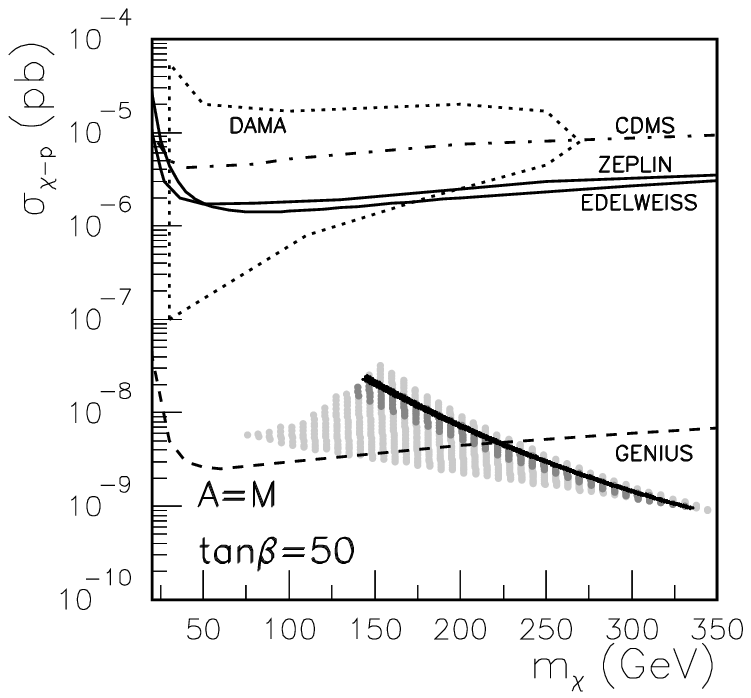,width=9cm}
%
\epsfig{file=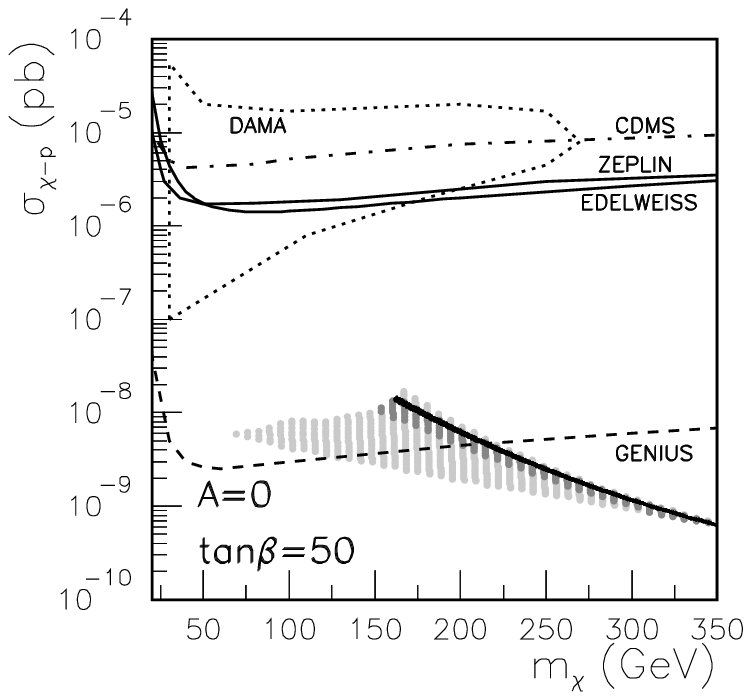,width=9cm}
\hspace*{-1.2cm}\epsfig{file=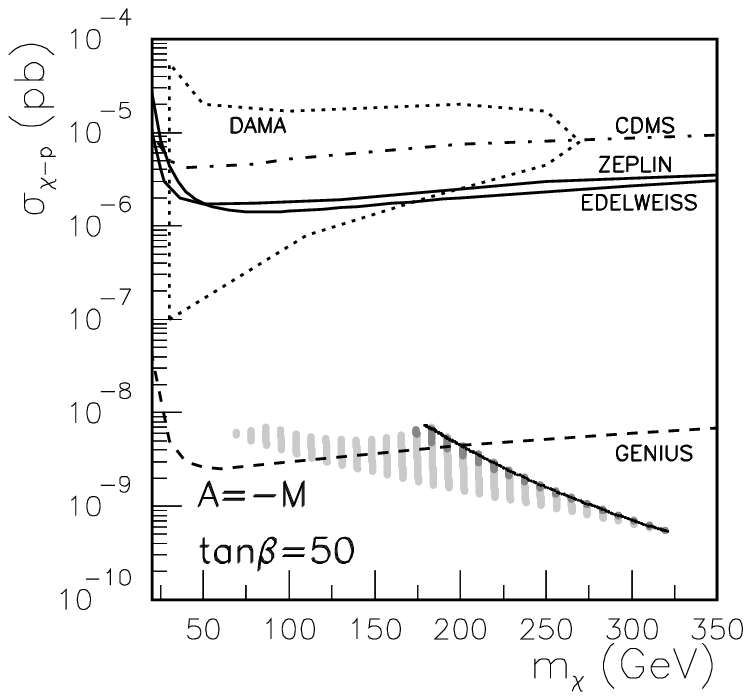,width=9cm}
%
\epsfig{file=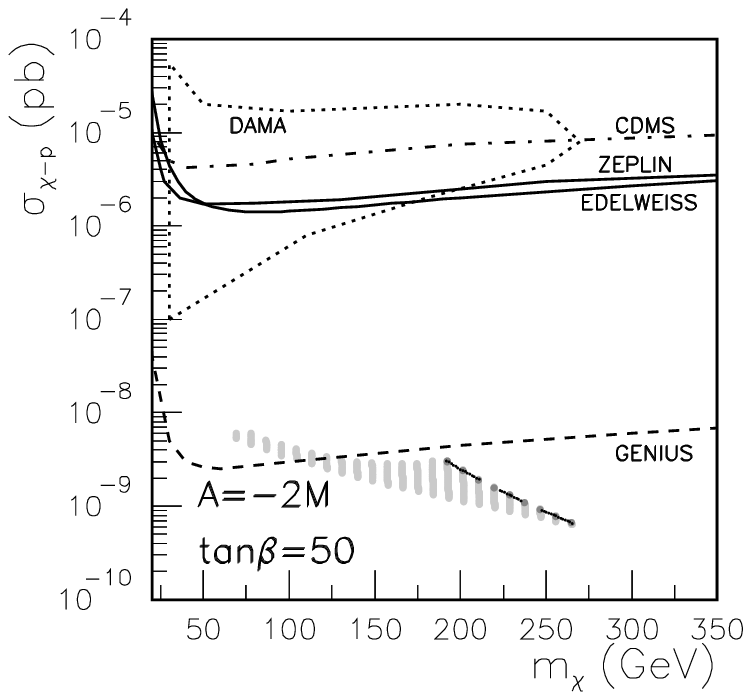,width=9cm}
\captions{The same as in Fig.~\ref{cross_scale122}
but for 
$\tan\beta=50$ and different values of $A$.
\label{cross_scale1226}}
\end{figure}

In Figs.~\ref{cross_scale122} and \ref{cross_scale1226} 
we summarize the above results for
$\tan\beta=35,50$, concerning the
cross section, showing the values of $\sigma_{\tilde{\chi}_{1}^{0}-p}$
allowed by all experimental constraints 
as a function of the neutralino mass
$m_{\tilde{\chi}_1^0}$, for different values of $A$.
Note that the gaugino mass
$M$ is essentially fixed for a given $\tan\beta$ and 
$m_{\tilde{\chi}_1^0}$.
Dark grey dots 
correspond to 
those points having a relic neutralino density within
the preferred range $0.1\leq\Omega h^2\leq 0.3$.
Given the narrow range of these points for the case 
$\tan\beta=35$,
they overlap 
in 


\clearpage

\thispagestyle{empty}

\clearpage

\clearpage


\clearpage

\clearpage

\noindent
the figure for $A=0$ with those
excluded by the UFB-3 constraint (shown with circles).
We observe that, generically, the cross section and the neutralino mass
are constrained to be
$5\times 10^{-10}\lsim \sigma_{\tilde{\chi}_1^0-p}\lsim 3\times 10^{-8}$ pb and
$120\lsim m_{\tilde{\chi}_1^0}\lsim 320$ GeV, respectively.
 
Obviously, in this mSUGRA case, more sensitive detectors
producing further data 
are needed\footnote{For a different conclusion, 
using a phenomenological SUSY model whose parameters are defined
directly at the electroweak scale, see ref.~\cite{effective}.}.
Fortunately, many dark matter detectors are being projected.
Particularly interesting is the case of GENIUS \cite{GENIUS},
where values of the cross section as low as 
$\approx 10^{-9}$ pb will be accessible, as shown in Figs.~4 and 5. 
It is worth noticing here that the DAMA area shown in the Figures,
has been obtained in ref.~\cite{halo} taking into account the
effect of the galactic halo modeling on the DAMA annual modulation
result. This area is larger than in the case 
of considering just the standard (isothermal sphere) model.
On the other hand, for the other experiments similar analyses
taking into account the uncertainties in the galactic halo 
are not available, and we use in the Figures the effect of the
standard halo model on their results. Including
these uncertainties, the region favoured by DAMA and not excluded
by the null searches would be in principle smaller than the one
shown here.

\subsection{Intermediate scale}


The analysis of the cross section
$\sigma_{\tilde{\chi}_1^0-p}$
carried out above in the context of mSUGRA,
was performed assuming the
unification scale
$M_{GUT} \approx 10^{16}$ GeV.
However, there are several interesting phenomenological 
arguments in favour of SUGRA scenarios
with scales 
$M_I\approx 10^{10-14}$ GeV (for a review see
e.g. ref.~\cite{darkcairo}).
In addition,
the string scale may be anywhere between the weak and the Planck 
scale, and explicit scenarios with
intermediate scales may arise in the context of 
D-brane constructions from type I strings \cite{nosotros}.
In this sense, to use the value of the initial scale,
say $M_I$, as a free parameter for the running of the soft terms
is particularly interesting.
In fact, it was  
pointed out in refs.~\cite{muas,Bailin,nosotros} 
that 
$\sigma_{\tilde{\chi}_1^0-p}$
is very sensitive to the variation of the initial scale 
for the running of the soft terms.
For instance, by taking $M_I=10^{10-12}$ GeV rather than 
$M_{GUT}$, and non-universal gauge couplings, 
regions in the parameter space of mSUGRA can be found 
where  $\sigma_{\tilde{\chi}_1^0-p}$ is two orders of magnitude
larger than for $M_{GUT}$ \cite{muas,nosopro}.

\begin{figure}
\hspace*{-1.2cm}\epsfig{file=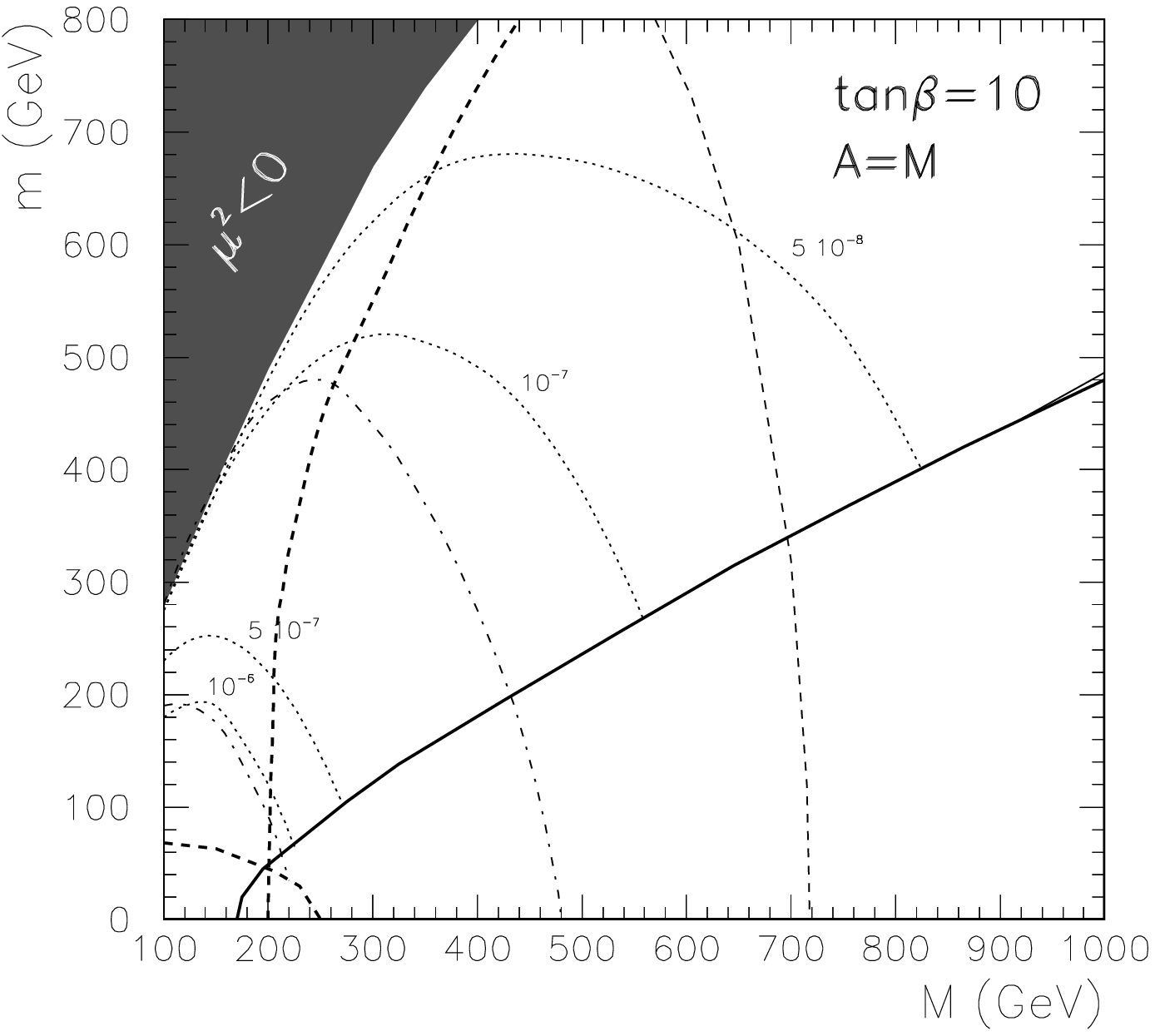,
width=9cm}
%
\epsfig{file=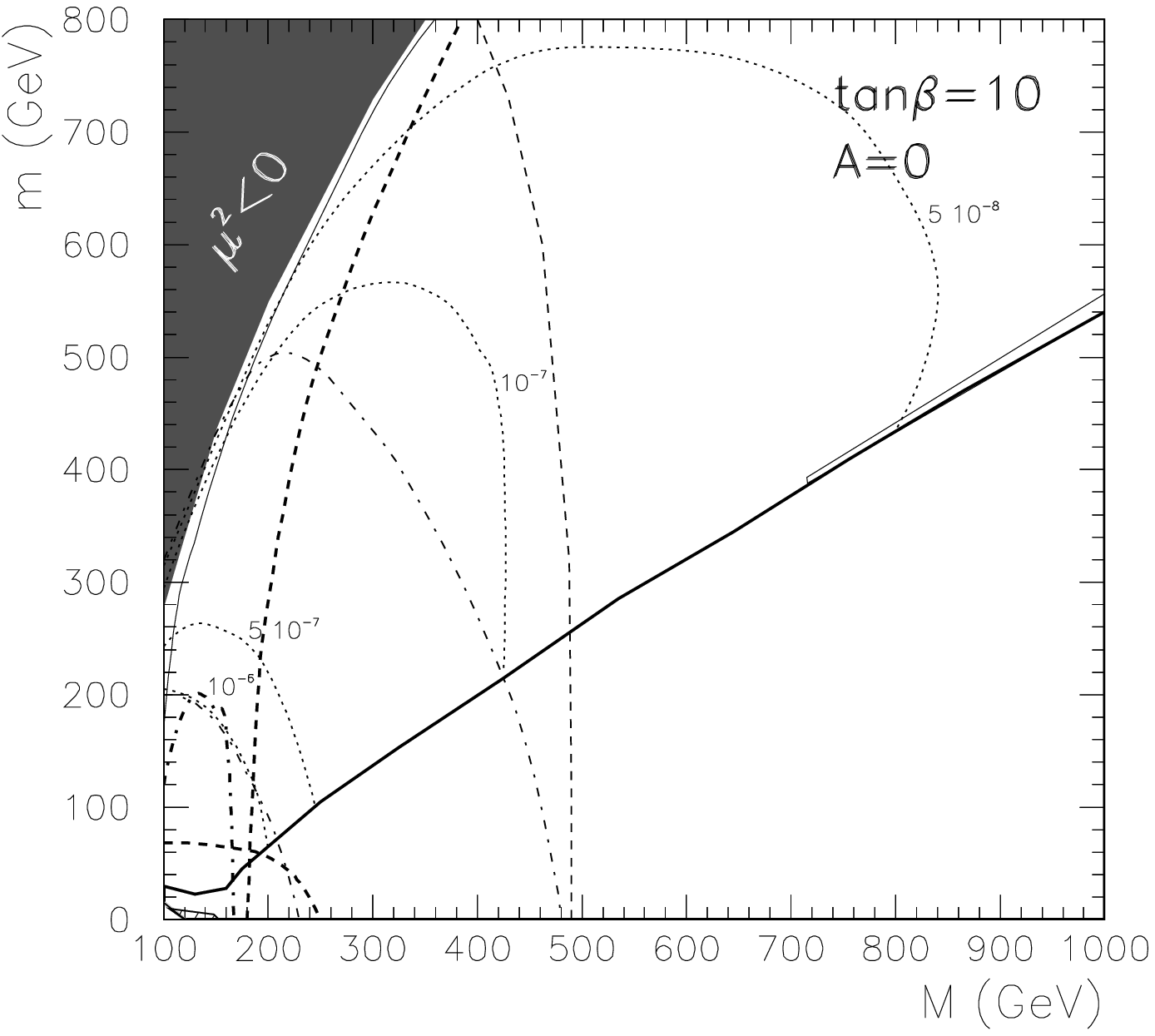,
width=9cm}

\hspace*{-1.2cm}\epsfig{file=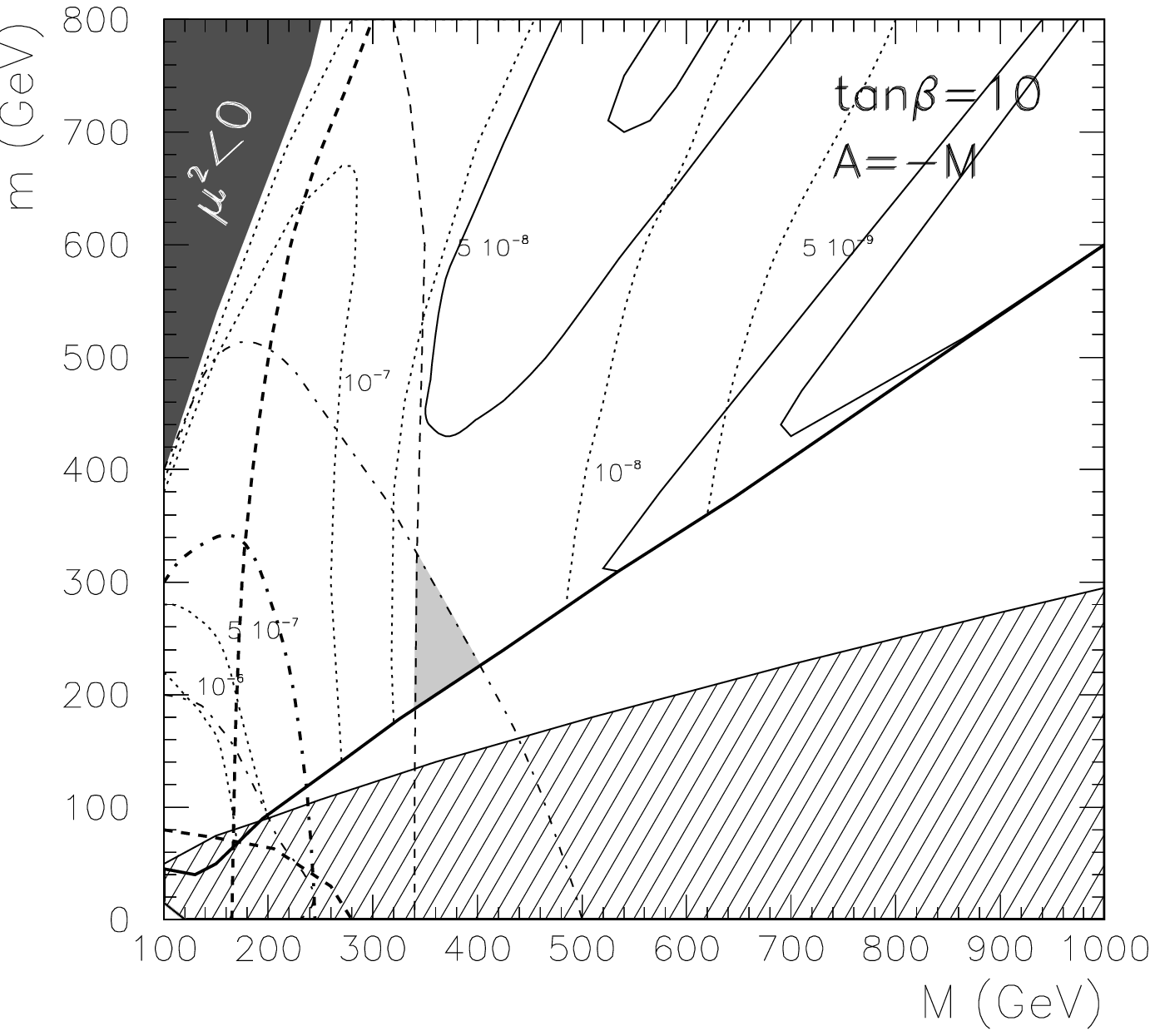,
width=9cm}
%
\epsfig{file=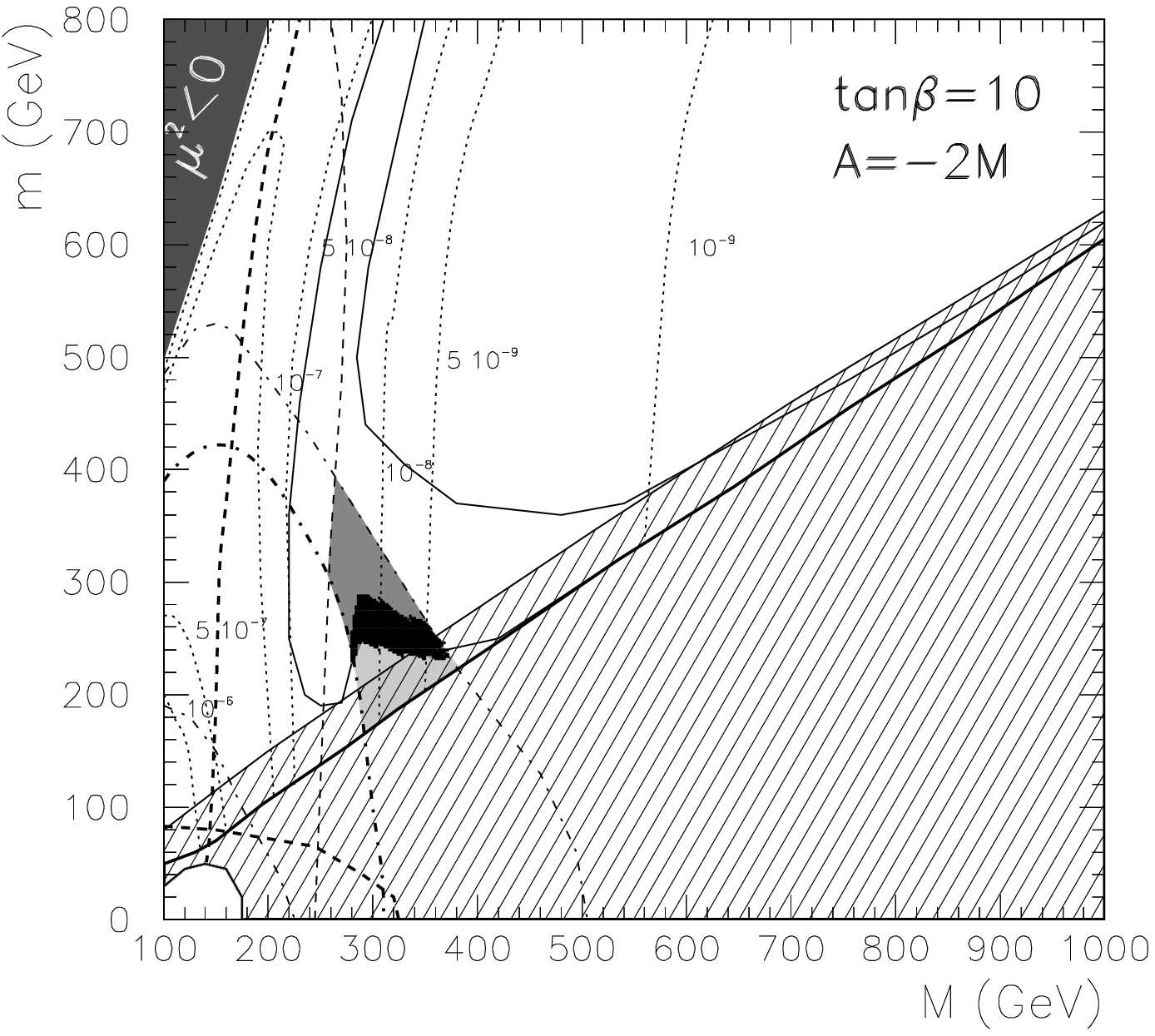,
width=9cm}
\captions{The same as in Fig.~\ref{a2m} but for the 
intermediate scale 
$M_I=10^{11}$ GeV, with $\tan\beta=10$ and different values of $A$. 
The black area is excluded because
$\mu^2$ becomes negative.
The white region at the bottom bounded by a solid line
is excluded because $m_{\tilde{\tau}_1}^2$ becomes negative.
\label{scale12_10}}
\end{figure}

\begin{figure}
\begin{center}
\epsfig{file=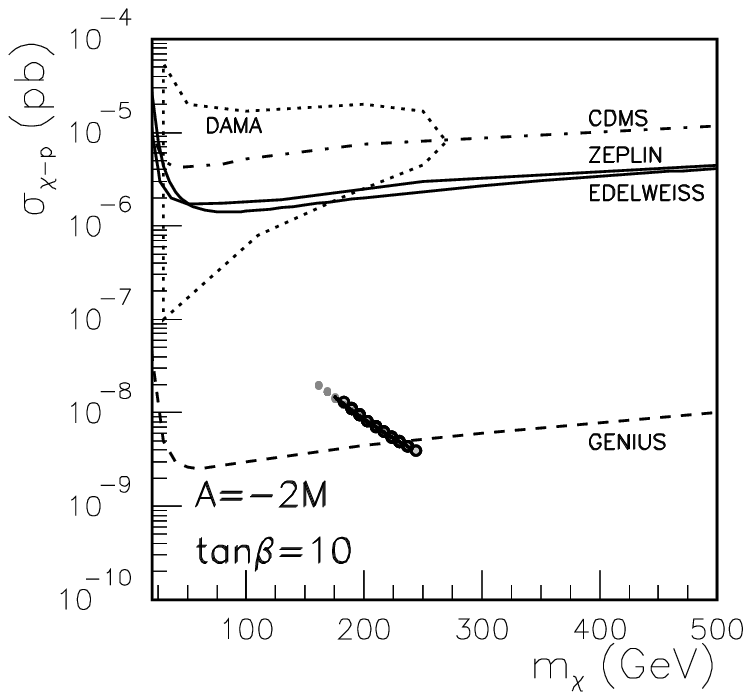,width=9cm}
\end{center}
\captions{The same as in Fig.~\ref{cross_scale122}
but for the intermediate scale
$M_I=10^{11}$ GeV,
with
$\tan\beta=10$ and $A=-2M$.
\label{cross_scale12}}
\end{figure}

The fact that smaller scales imply a larger 
$\sigma_{\tilde{\chi}_1^0-p}$  
can be explained with the variation in the value of 
$\mu$ with $M_I$.
One observes that, for $\tan\beta$ fixed, the smaller the initial
scale for the running the smaller the numerator in the
first piece of eq.~(\ref{electroweak}) becomes. 
This can be understood from 
the well known evolution of $m_{H_u}^2$ with the
scale.
Clearly, when the value of the initial scale is reduced 
the RGE running is shorter and, as a consequence, 
the negative contribution 
$m_{H_u}^2$ to $\mu^2$ in eq.~(\ref{electroweak}) becomes less important.
Then, 
$|\mu|$ decreases and therefore
the Higgsino composition of the lightest neutralino increases.
Eventually, $|\mu|$ will be of the order of $M_1$, $M_2$
and $\tilde{\chi}_1^0$ will be a mixed Higgsino-gaugino state.
In addition, when $|\mu|$ decreases the (tree-level)
mass of the CP-odd Higgs,
$m^2_A=m_{H_d}^2+m_{H_u}^2+2\mu^2$ decreases.
Since the heaviest CP-even Higgs,
$H$, is almost degenerate in mass with this, it also decreases
significantly.
Indeed, scattering channels through Higgs exchange 
are very important and their contributions to the cross section
will increase it.
Let us also remark that, for the same value of the parameters, 
the Higgs mass $m_h$ decreases with respect to the GUT scale scenario.
This is because the value of $m_h$ depends on the value of
the gluino mass $M_3$. It increases when $M_3$ increases at low energy.
However, now the running is shorter and therefore $M_3$ at low energy
is smaller than in the GUT scenario.
Although the latter may be welcome in order to obtain larger
cross sections, it may also be dangerous when
confronting with the experimental result concerning 
the Higgs mass.

Concerning the value of the relic density,
$\Omega_{\tilde{\chi}_1^0}$ is dramatically reduced
with respect to the $M_{GUT}$ case. 
This is due to a combination of several factors:
1) The Higgsino-gaugino composition of $\tilde{\chi}_1^0$
allows a significant increase of the $\tilde{\chi}_1^0$ annihilation
cross section, due to channels with Higgs and gauge bosons
in the final states; 2) The decrease of the mass of the pseudoscalar
Higgs, along with the value of the $\mu$--term, enables the presence of
resonant annihilation channels even at $\tan\beta=10$;
3) The masses of the lightest chargino and stop are small enough
to allow $\tilde{\chi}_1^0$--$\tilde{\chi}_1^\pm$ \cite{char} 
and $\tilde{\chi}_1^0$--$\tilde{t}_1$ \cite{stop}
coannihilations in some areas of the parameter space. Although 
the later is less relevant, we find some areas at $\tan\beta=50$ and $A<0$.

We show in 
Fig.~\ref{scale12_10} the result for $M_I=10^{11}$ GeV, with
$\tan\beta=10$.
This can be compared with the one in
Fig.~\ref{a2m}, where $M_{GUT}$ is used. 
Now the relation $\neumass\sim 0.4\ M$ 
does not hold, and one has $\neumass> 0.4M$. In any case, 
$\neumass<M_1$ since the bino-Higgsino mixing is significant in
this case.
Clearly, for the same values of the parameters,
larger cross sections can be obtained with the intermediate scale.
It is worth noticing that even with this moderate value of 
$\tan\beta$, $\tan\beta=10$,
there are regions where the cross

\begin{figure}
\begin{center}
\hspace{1.5cm}\epsfig{file=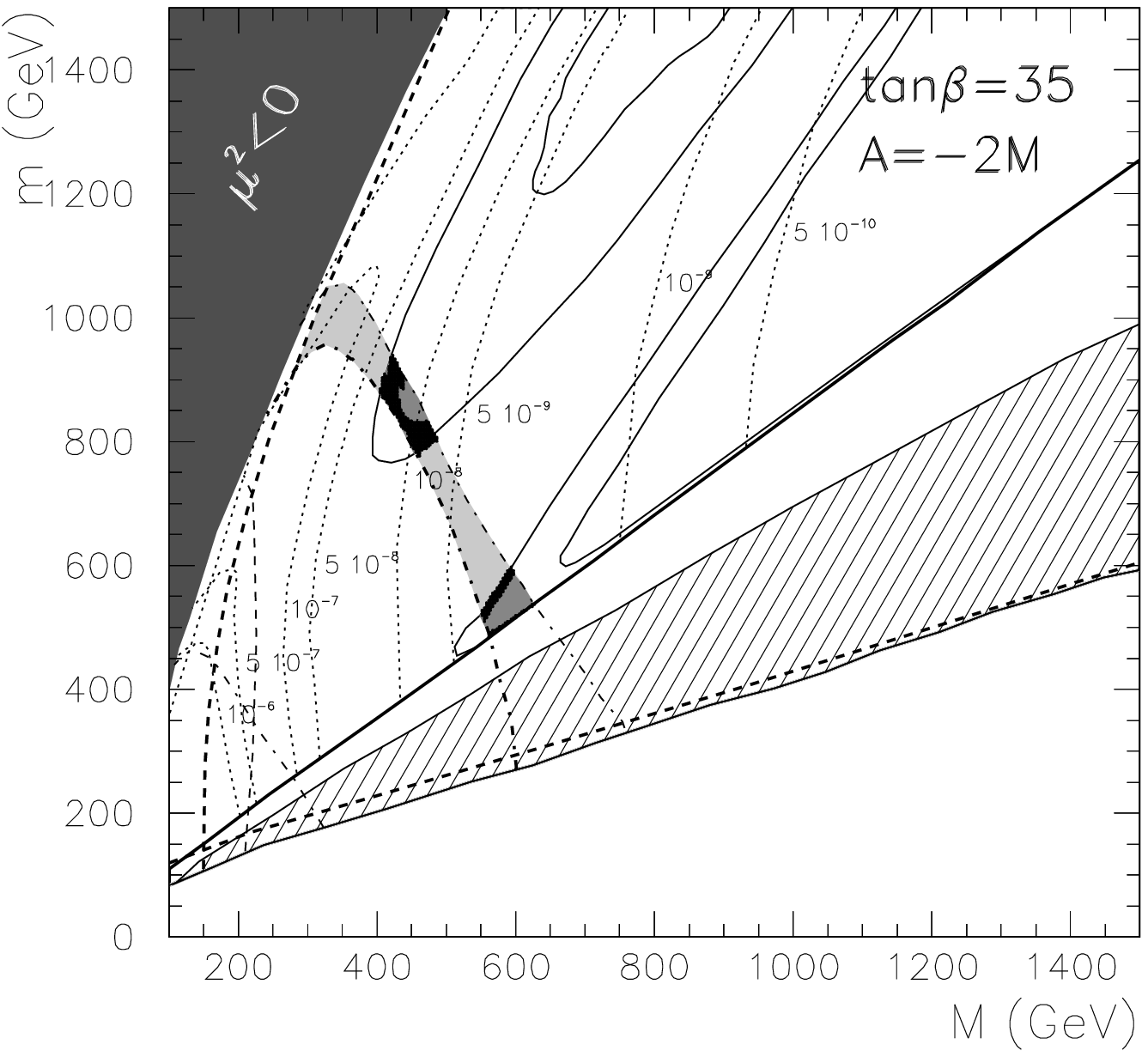,
width=9cm}
\end{center}
\captions{The same as in Fig.~6 
but for 
$\tan\beta=35$ and $A=-2M$.
\label{scale12_106}}
\end{figure}

\begin{figure}
\begin{center}
\epsfig{file=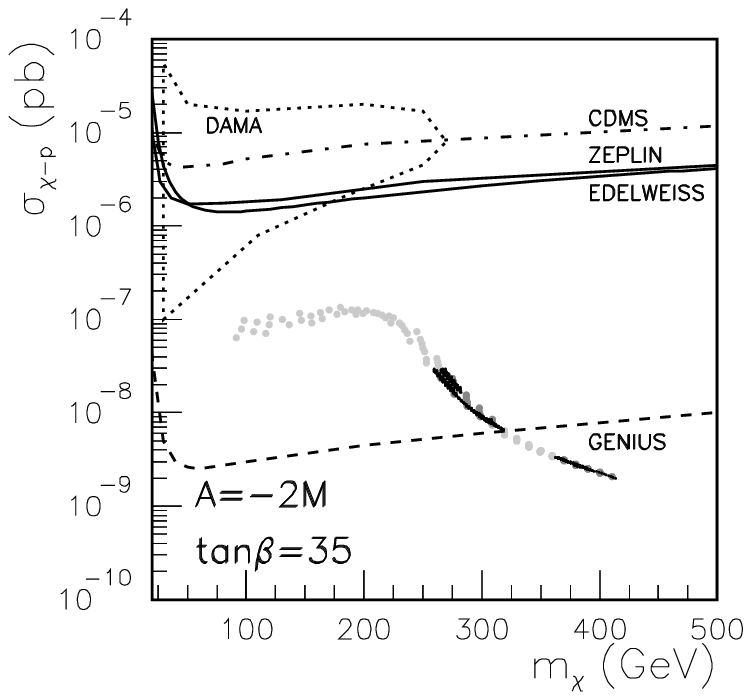,width=9cm}
\end{center}
\captions{The same as in Fig.~7
but for 
$\tan\beta=35$ and $A=-2M$.
\label{cross_scale1221}}
\end{figure}

\begin{figure}
\hspace*{-0.7cm}\epsfig{file=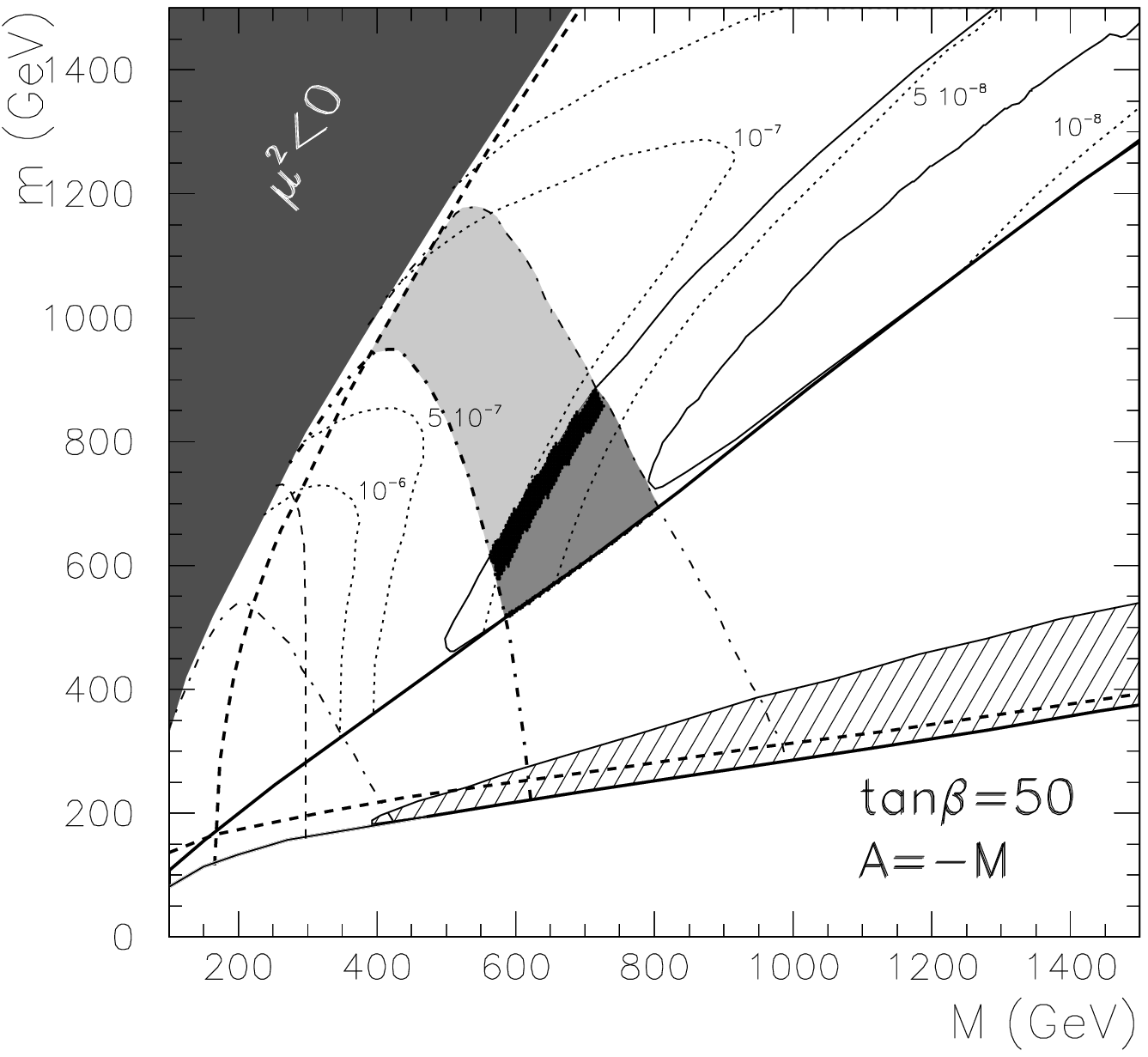,width=9cm}\,\,\,\,\,\,\,\,\,\,\,
\hspace*{-0.7cm}\epsfig{file=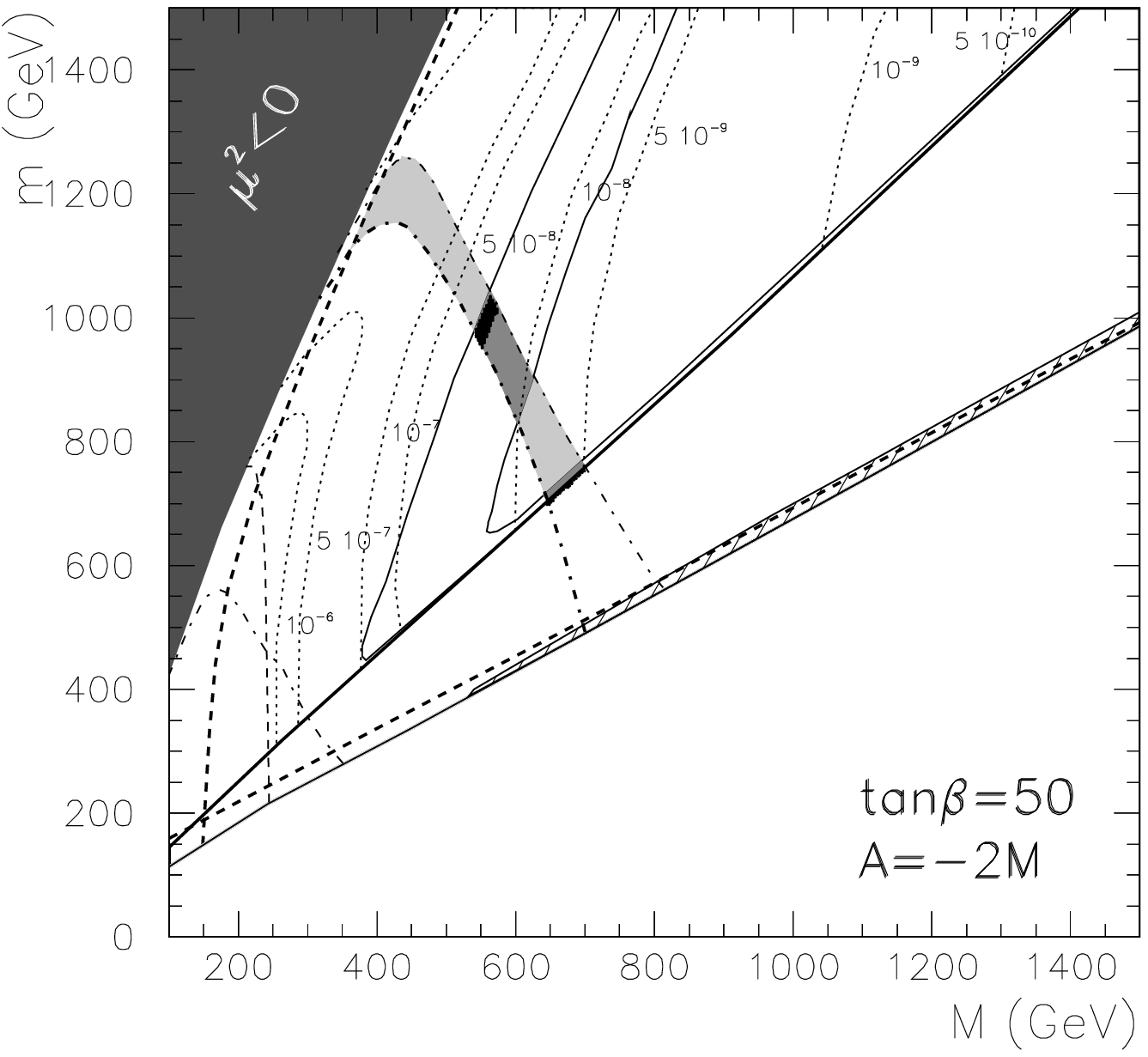,width=9cm}
\captions{The same as in Fig.~6
but for 
$\tan\beta=50$ and $A=-M,-2M$.
\label{scale12_1066}}
\end{figure}

\begin{figure}
\hspace*{-1.2cm}\epsfig{file=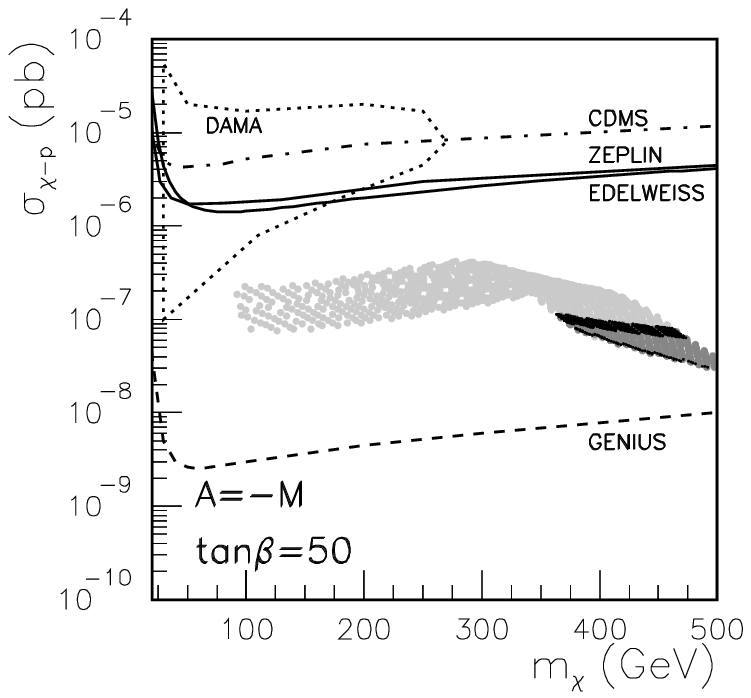,
width=9cm}\,\,\,\,\,\,\,\,\,\,\,
\hspace*{-1.2cm}\epsfig{file=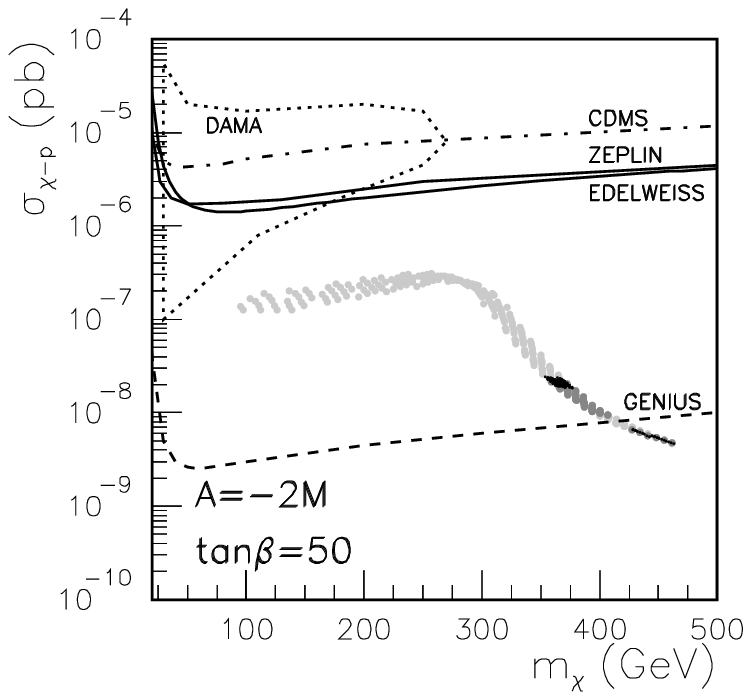,width=9cm}
\captions{The same as in Fig.~7
but for 
$\tan\beta=50$ and $A=-M,-2M$.
\label{cross_scale1222}}
\end{figure}

\noindent section enters in the DAMA 
area, 
$\sigma_{\tilde{\chi}_1^0-p}\approx 10^{-6}$ pb.
However, 
for $A=M,0$
the whole parameter space is forbidden
due to the combination of the Higgs mass bound with
the $g_{\mu}-2$ lower bound.
We have checked explicitly that 114.1 GeV is the correct lower bound to be
used
concerning the Higgs mass, since generically
$\sin^2(\alpha-\beta)\sim 1$ for the intermediate scale.
Notice also that now, for these values of $A$,
$\Omega_{\tilde{\chi}_1^0}h^2$ is smaller than 0.1 in most of the
parameter space. Only very small regions bounded by solid lines,
and therefore with 
$0.1\leq \Omega_{\tilde{\chi}_1^0}h^2\leq 0.3$,
can be found in both figures with $A=M,0$.
For $A=-M,-2M$ there are small regions where the
$m_h$ and $g_{\mu}-2$ bounds
are compatible, but finally the constraint $m_h>114.1$ GeV implies
that the allowed cross sections do not enter in the DAMA area.
Although now larger regions with 
$0.1\leq \Omega_{\tilde{\chi}_1^0}h^2\leq 0.3$
are present, for $A=-M$ these are not compatible with the
experimental bounds.

We also find that the region excluded by the UFB-3 constraint 
is much smaller than in those cases where the initial 
scale is the GUT one (see Fig.\,\ref{a2m}).
In fact, only for $A=-2M$ this region is larger than the one
forbidden by the LSP bound.
As explained above, 
when the initial scale is smaller the
negative value of 
$m_{H_u}^2$ becomes less important. Therefore 
the negative contribution to $V_{UFB-3}$
is also less important.
As a consequence, the UFB-3 minima are not so easily
deeper than the realistic one \cite{another} 
\footnote{This argument was used in ref.~\cite{Abel} to allow again the
dilaton dominated supersymmetry breaking scenario, which
was essentially ruled out by the UFB-3 constraint for the heterotic string 
scale \cite{clm2}.}.

In Fig.~\ref{cross_scale12} we summarize the above results for
$\tan\beta=10$, concerning the
cross section, showing the values of $\sigma_{\tilde{\chi}_{1}^{0}-p}$
allowed by all experimental constraints 
as a function of the neutralino mass
$m_{\tilde{\chi}_1^0}$, for $A=-2M$.
Only in this case there are dark grey dots corresponding
to points having a relic neutralino density within the
preferred range
$0.1\leq \Omega_{\tilde{\chi}_1^0}h^2\leq 0.3$.
Given the narrow range of these points,
they overlap in the figure with those
excluded by the UFB-3 constraint.
They correspond to
$150\lsim m_{\tilde{\chi}_1^0}\lsim 250$ GeV, and e.g. the mass
of the lightest stop, $\tilde{t_1}$, is between 250 and 340 GeV. 

Qualitatively, similar results are obtained for larger values of $\tan\beta$.
For example, for $\tan\beta=35$ and $A=M$ there are now regions in the
parameter space with $\sigma_{\tilde{\chi}_{1}^{0}-p}\lsim 3\times 10^{-7}$ pb
fulfilling all experimental constraints, however they have
$\Omega_{\tilde{\chi}_1^0}h^2< 0.03$.
A similar situation occurs for $A=0$, where  
$\Omega_{\tilde{\chi}_1^0}h^2< 0.06$. 
On the other hand, for $A=-2M$ we obtain
points allowed by all experimental and astrophysical constraints,
and this is shown in Figs.~8 and 9. 
The upper bound in the cross section 
is because of the $b\to s\gamma$ process.
Points within the preferred astrophysical range correspond to 
$\sigma_{\tilde{\chi}_{1}^{0}-p}\lsim 10^{-8}$ pb.
Now, there are two allowed regions with 
$275\lsim m_{\tilde{\chi}_1^0}\lsim 325$ GeV and
$370\lsim m_{\tilde{\chi}_1^0}\lsim 420$ GeV. For these, 
$570\lsim m_{\tilde{t_1}}\lsim 720$ GeV.

The situation for $\tan\beta=50$ is shown in Figs.~10 and 11.
Now, unlike the previous case, 
for $A=-M$ there are also points allowed by all experimental and
astrophysical
constraints, and 
$\sigma_{\tilde{\chi}_{1}^{0}-p}\lsim 10^{-7}$ pb. 
In this case, for example for $A=-2M$,
there are two allowed regions with
$360\lsim m_{\tilde{\chi}_1^0}\lsim 410$ GeV,
$600\lsim m_{\tilde{t_1}}\lsim 800$ GeV
and 
$425\lsim m_{\tilde{\chi}_1^0}\lsim 460$ GeV,
$445\lsim m_{\tilde{t_1}}\lsim 475$ GeV.
We should mention that in the
later the effect of ${\tilde{\chi}_1^0}$--${\tilde{t_1}}$
is present in the estimation of $\Omega_{\tilde{\chi}_1^0}h^2$, however
it is less significant than
the ${\tilde{\chi}_1^0}$--${\tilde{\tau_1}}$ one.

\section{SUGRA scenario with non-universal soft terms}

 
As mentioned in the Introduction,
the general situation for 
the soft parameters is to have a
non-universal structure, and in fact, generic string constructions, 
whose low-energy limit is SUGRA,
exhibit these properties.
It was shown in the literature
that the non-universality of the soft parameters allows to 
increase 
the neutralino-proton cross section. 
This can be carried out with non-universal 
scalar masses \cite{Bottino,Arnowitt,Drees,darkcairo,Santoso,nosopro,Dutta}
and/or gaugino masses \cite{Nath2,darkcairo,nosopro,Orloff,Dutta,Birkedal}.
We will concentrate on this possibility here, taking into account the
effect of the charge and colour breaking constraints.

\vspace{0.5cm}

\noindent {\it (i) Non-universal scalar masses}

\vspace{0.5cm}

Let us analyse for the moment a SUGRA scenario with GUT scale
and non-universal soft scalar masses.
This non-universality can be parameterized in the Higgs sector as
follows:
\begin{eqnarray}
m_{H_{d}}^2=m^{2}(1+\delta_{1})\ , \quad m_{H_{u}}^{\ 2}=m^{2}
(1+ \delta_{2})\ .
\label{Higgsespara}
\end{eqnarray}
Concerning squarks and sleptons, in order to avoid potential problems
with
flavour changing neutral currents, one can assume that the first
two generations have a common scalar mass $m$ at $M_{GUT}$, and
that non-universalities are allowed only for the third generation:
\begin{eqnarray}
m_{Q_{L}}^2&=&m^{2}(1+\delta_{3})\ , \quad m_{u_{R}}^{\ 2}=m^{2}
(1+\delta_{4})\ , 
\nonumber\\
m_{e_{R}}^2&=&m^{2}(1+\delta_{5})\ ,  \quad m_{d_{R}}^{\ 2}=m^{2}
(1+\delta_{6})\ , \nonumber\\
m_{L_{L}}^2&=&m^{2}(1+\delta_{7})\ ,     
\label{Higgsespara2}
\end{eqnarray}
where  
$Q_{L}=(\tilde{t}_{L}, \tilde{b}_{L})$, $L_{L}=(\tilde{\nu}_{L},
\tilde{\tau}_{L})$, $u_{R}=\tilde{t}_{R}$ and $e_{R}=\tilde{\tau}_{R}$.
Note that whereas $\delta_{i} \geq -1 $, $i=3,...,7$, in order to avoid
an UFB direction breaking charge and colour, 
$\delta_{1,2} \leq -1$
is possible as long as the conditions 
$m_1^2=m_{H_{d}}^2+\mu^2>0$, $m_2^2=m_{H_{u}}^2+\mu^2>0$  
are fulfiled. 

As discussed for intermediate scales in Subsection~4.2,
an important factor in order to increase the cross section, 
consists in reducing the value of $|\mu|$.
This value is determined by condition (\ref{electroweak}) and can
be significantly reduced for some choices of the $\delta$'s. We can have
a qualitative
understanding of the effects of the $\delta$'s on $\mu$ from
the following.
First, when $m_{H_u}^2$ at $M_{GUT}$ increases
its negative low-energy contribution to eq.~(\ref{electroweak})
becomes less important. Second, when $m_{Q_{L}}^2$ and $m_{u_{R}}^2$
at $M_{GUT}$ decrease, due to their contribution proportional
to the top Yukawa coupling in the RGE of $m_{H_u}^2$, the negative
contribution of the latter to $\mu^2$ is again less important.
Thus one can deduce that 
$\mu^2$ will be reduced (and hence $\sigma_{\tilde{\chi}_1^0-p}$ increased)
by choosing $\delta_{3,4} < 0$ and $\delta_2 >0$.
In fact non-universalities in the Higgs sector give the most important
effect, and including the one in the sfermion sector the cross
section only increases slightly. Thus in what follows we will take
$\delta_{i}=0$, $i=3,...,7$.

Concerning the value of the relic density,
$\Omega_{\tilde{\chi}_1^0}$ is affected due to the increase
of the Higgsino  components of $\tilde{\chi}_1^0$ with respect
to
the dominant bino component of the universal case. The change in $\mu$ also
determines the presence of the Higgs mediated resonant channels. 
In contrast
to Subsection~4.2, the most relevant coannihilation
scenarios are ${\tilde{\chi}_1^0}$--${\tilde{\tau_1}}$, in particular
$\tilde{\chi}_1^0$--$\tilde{\chi}_1^\pm$ coannihilations are only sizeable
for $\tan\beta=35$ in Fig.~12 (see the discussion below), 
when the $\mu$--parameter becomes small.
However, even in this case the area inside the WMAP bounds corresponds to
neutralino annihilations, which are enhanced due to enlargement of its
Higgsino components.

On the other hand, there is another relevant 
way of increasing the cross section using the non-universalities
of the Higgs sector. Note that 
decreasing 
$m_{H_d}^2$, i.e. choosing $\delta_1 < 0$,
leads to a decrease in 
$m^2_A=m_{H_d}^2+m_{H_u}^2+2\mu^2$
and therefore in the mass of the heaviest 
Higgs\footnote{On the contrary, the lightest
Higgs mass, $m_h$, is almost unaltered, it only decreases less than 1\%.} 
$H$.
This produces an increase in the
cross section.

Thus we will see that,
unlike the universal scenario in Section~4, 
with non-universalities is possible
to obtain large
values of the cross section, and even some points 
enter in the DAMA area fulfilling
all constraints. 
Let us analyse three representative cases with
\begin{eqnarray}
a)\,\, \delta_{1}&=&0\ \,\,\,\,\,\,\,\,,\,\,\,\, \delta_2\ =\ 1\ ,
\nonumber\\
b)\,\, \delta_{1}&=&-1\ \,\,\,\, ,\,\,\,\, \delta_2\ =\ 0\ ,
\nonumber\\
c)\,\, \delta_{1}&=&-1\ \,\,\,\, ,\,\,\, \delta_2\ =\ 1 
\ .
\label{3cases}
\end{eqnarray}

Clearly, the above discussion about decreasing $\mu^2$ applies
well to case {\it a)}, where the variation in
$m_{H_u}^2$ through $\delta_2$ is relevant.
This is shown in Fig.~12 for $\tan\beta=35,50$ and $A=0$, which
can be compared with 
Figs.~2 and 3.
Note that now, for $\tan\beta=35$, 
there is an important area in the upper left
where $\mu^2$ becomes negative due to the increasing in
$\delta_2$ with respect to the universal case.
A larger area is forbidden for large values of $\tan\beta$, as e.g.
$\tan\beta=50$, 
but now because $m_A^2$ becomes negative. This is similar to what occurs
in the universal scenario when $\tan\beta\gsim 60$, as mentioned
in Section~3.
Notice that eq.~(\ref{electroweak}) can be written as
$\mu^2\approx -m_{H_u}^2-\frac{1}{2}M_Z^2$ and therefore
$m^2_A\approx m_{H_d}^2-m_{H_u}^2-M_Z^2$.
Since
$m_{H_u}^2$ at $M_{GUT}$ increases
its negative low-energy contribution 
becomes less important. In addition,  
the bottom Yukawa coupling is large and the $m_{H_d}^2$
becomes negative. 
As a consequence $m_A^2<0$.

For $\tan\beta=35$,
although the cross section increases with respect to the universal
case, the present experimental constraints exclude points
entering in the DAMA area. 
This can be seen more clearly comparing Figs.~13 and 4.
Notice also that the astrophysical bounds
$0.1\lsim\Omega_{\tilde{\chi}_1^0}h^2\lsim 0.3$ imply
$\sigma_{\tilde{\chi}_1^0-p}\approx 10^{-8}$ pb.
On the contrary, for $\tan\beta=50$ there are points
entering in the DAMA area, and even part of them            
fulfil the astrophysical bounds.
We have checked that for $A=M$ the figures are similar, although
no points enter in the DAMA area, even for $\tan\beta=50$.
On the other hand, the region forbidden by
the LSP bound is larger than the one forbidden by the UFB-3 constraint.

We have also checked that larger values of $\delta_2$,
as e.g. $\delta_2=1.5$, give rise to similar figures.
For small values, $\delta_2\gsim 0.2$ is sufficient to enter in DAMA
fulfilling the
experimental bounds
with $\tan\beta=50$. In fact, e.g., for $\delta_2=0.5, 0.75$ one also gets
many points entering in DAMA as for $\delta_2=1$, however, they do not
fulfil the astrophysical bounds. For the latter one needs
$\delta_2>0.85$.

\begin{figure}
\hspace*{-0.7cm}\epsfig{file=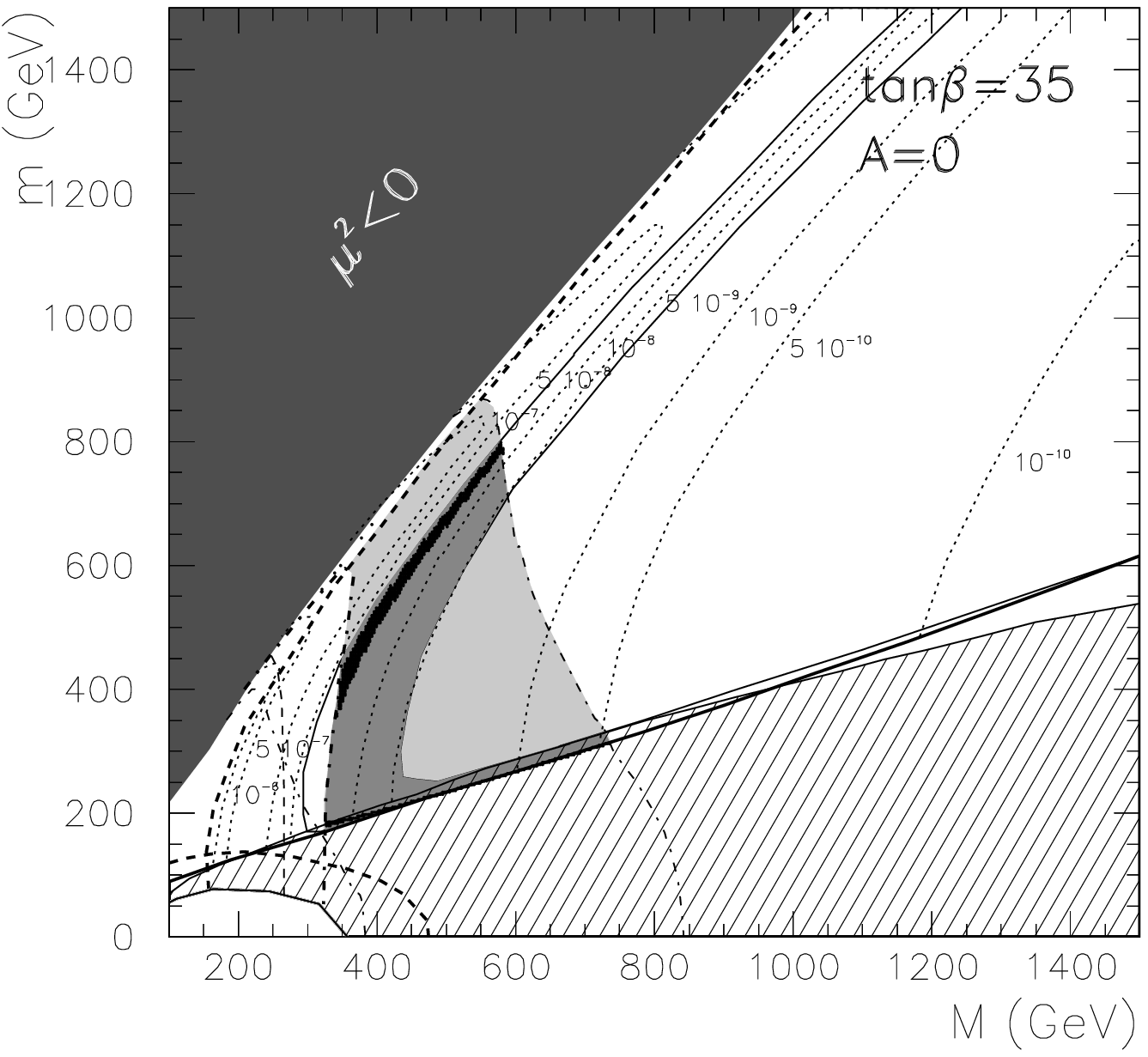,width=9cm}\,\,\,\,\,\,\,\,\,\,\,
%
\hspace*{-0.7cm}\epsfig{file=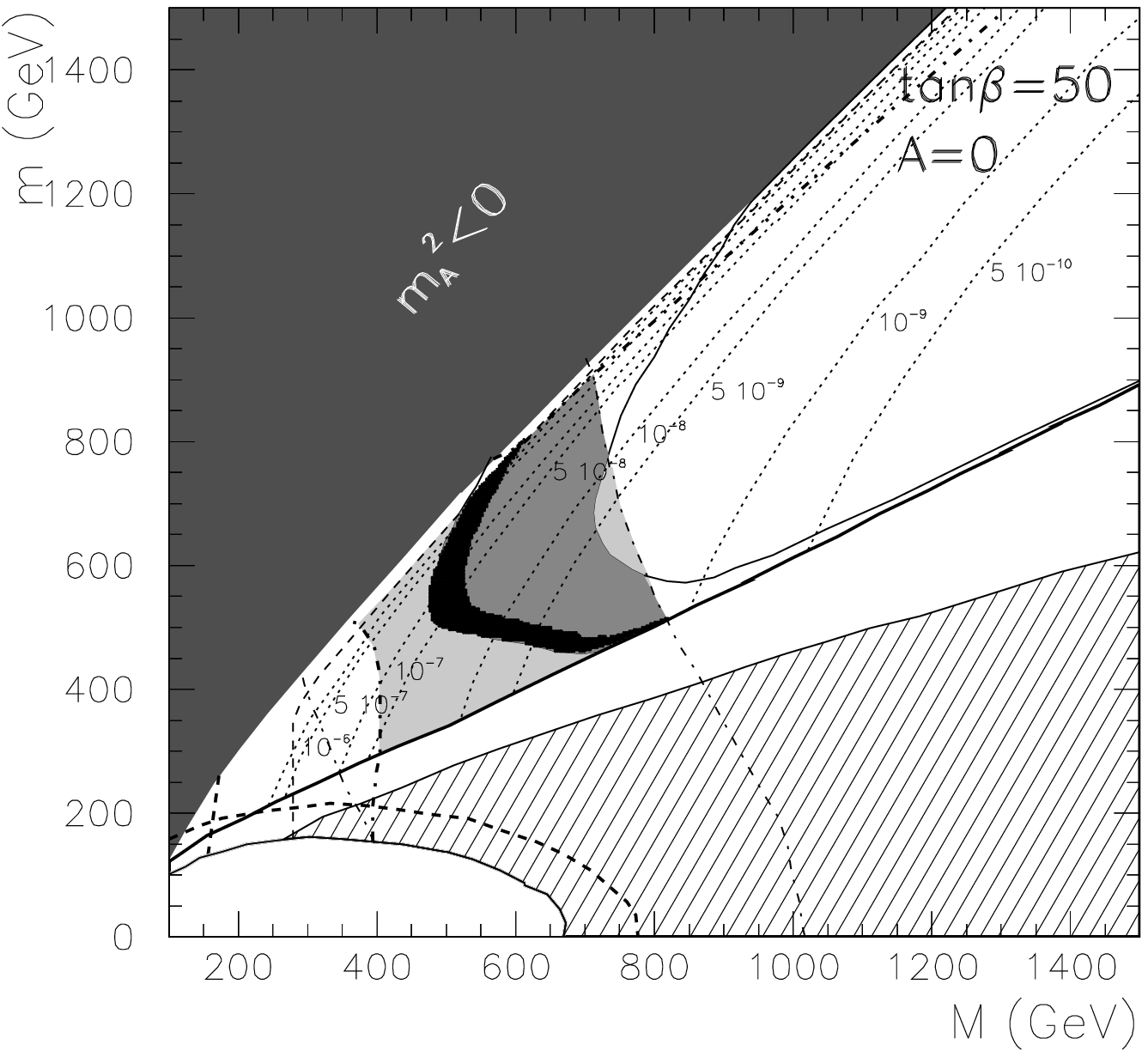,width=9cm} 



\captions{The same as in 
Figs.~2 and 3 
but for the non-universal case {\it a)} $\delta_1=0$, $\delta_2=1$, 
discussed in eq.~(\ref{3cases}), with $\tgb=35,50$ and $A=0$.
\label{anu}}
\end{figure}

\begin{figure}
\hspace*{-1.2cm}\epsfig{file=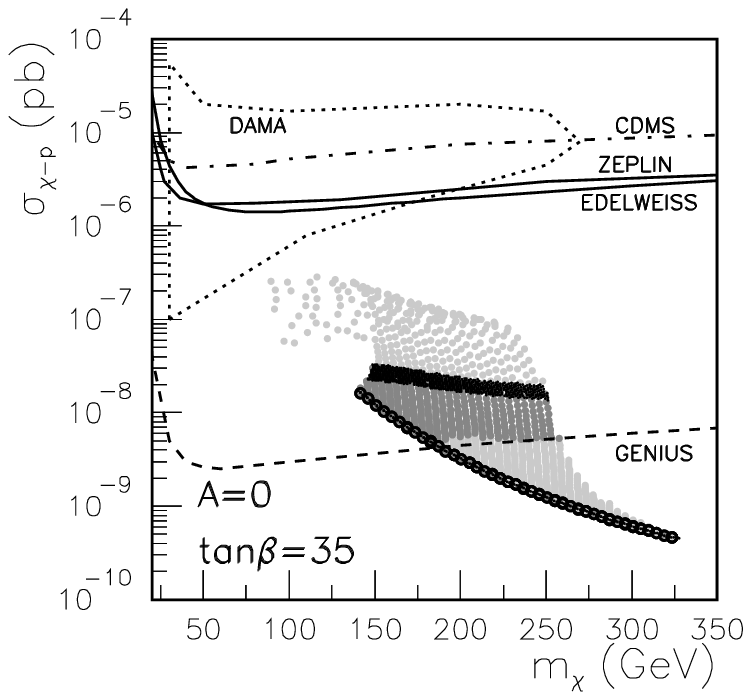,width=9cm}\,\,\,\,\,\,\,\,\,\,\,
%
\hspace*{-1.2cm}\epsfig{file=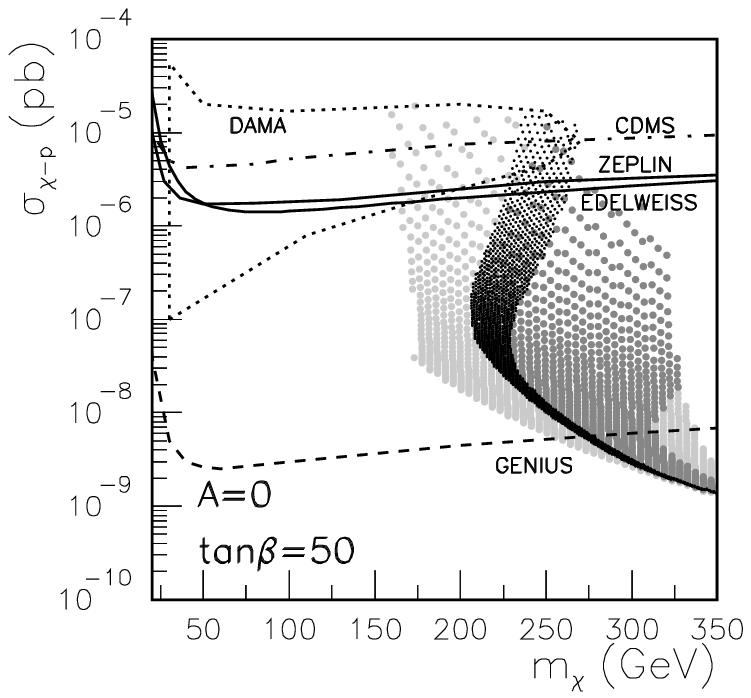,width=9cm}


\captions{The same as in Figs.~\ref{cross_scale122}
and 5 but for 
the non-universal case {\it a)} $\delta_1=0$, $\delta_2=1$, 
discussed in eq.~(\ref{3cases}), with $\tgb=35,50$ and $A=0$.
\label{anusec}}
\end{figure}

\begin{figure}
\hspace*{-0.7cm}\epsfig{file=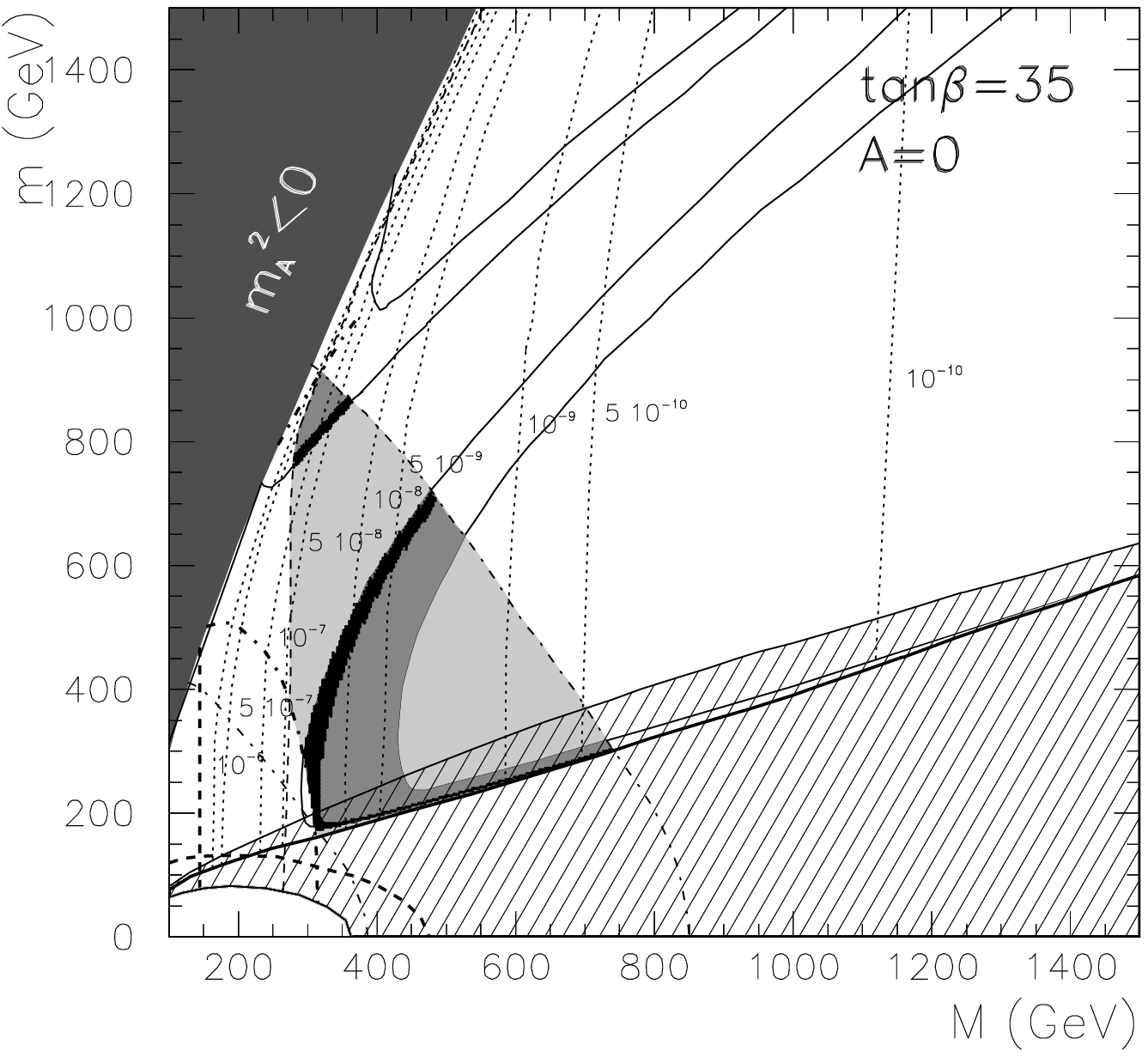,width=9cm}\,\,\,\,\,\,\,\,\,\,\,
%
\hspace*{-0.7cm}\epsfig{file=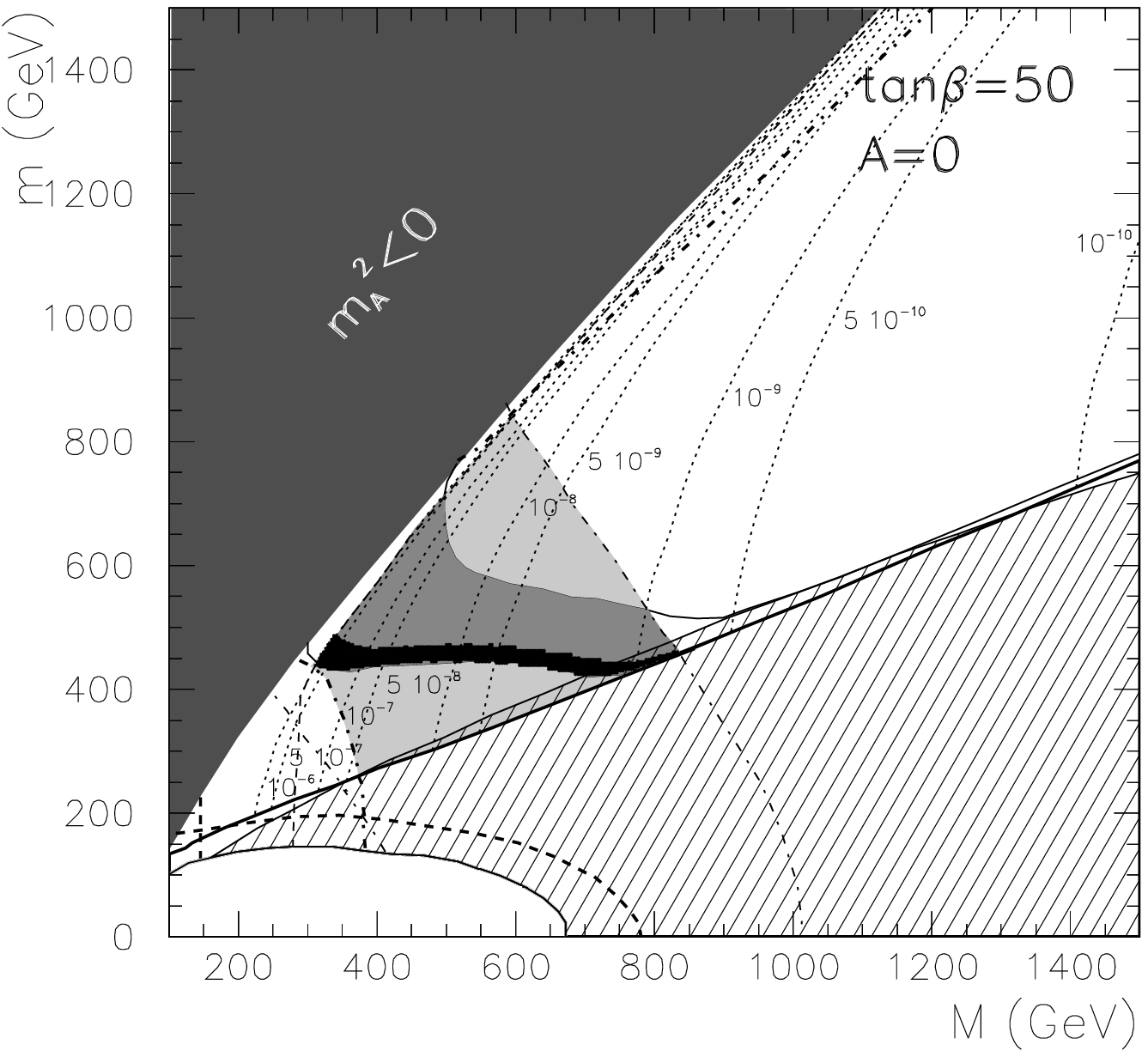,width=9cm} 



\captions{The same as in 
Figs.~2 and 3 
but for the non-universal case {\it b)} $\delta_1=-1$, $\delta_2=0$, 
discussed in eq.~(\ref{3cases}), with $\tgb=35,50$ and $A=0$.
\label{anu2}}
\end{figure}

\begin{figure}
\hspace*{-1.2cm}\epsfig{file=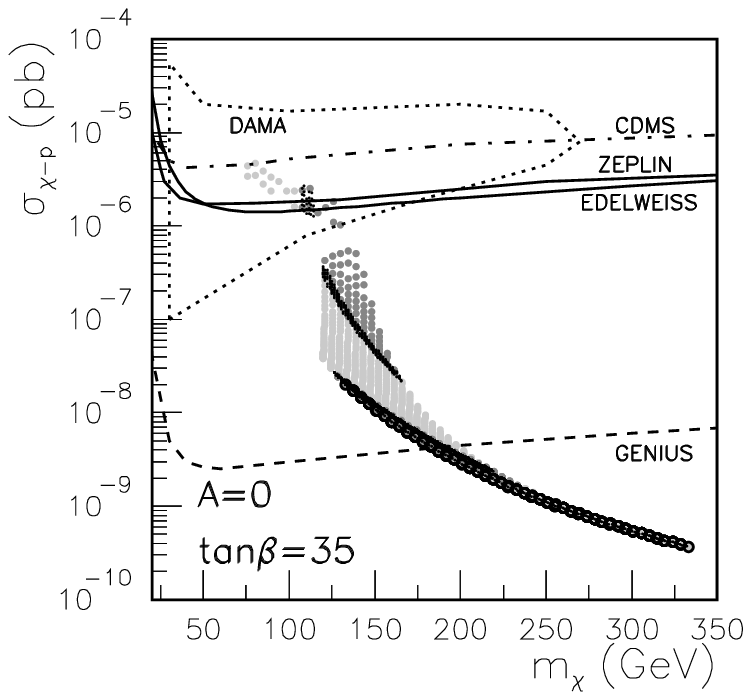,width=9cm}\,\,\,\,\,\,\,\,\,\,\,
%
\hspace*{-1.2cm}\epsfig{file=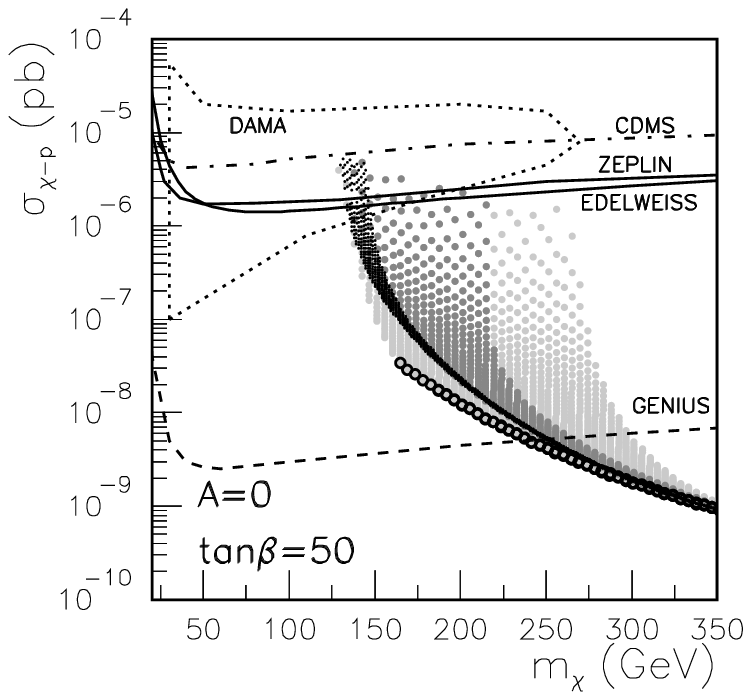,width=9cm}


\captions{The same as in Figs.~\ref{cross_scale122} and 5
but for the non-universal case {\it b)} $\delta_1=-1$, $\delta_2=0$, 
discussed in eq.~(\ref{3cases}), with $\tgb=35,50$ and $A=0$.
\label{anusec2}}
\end{figure}

\begin{figure}
\hspace*{-0.7cm}\epsfig{file=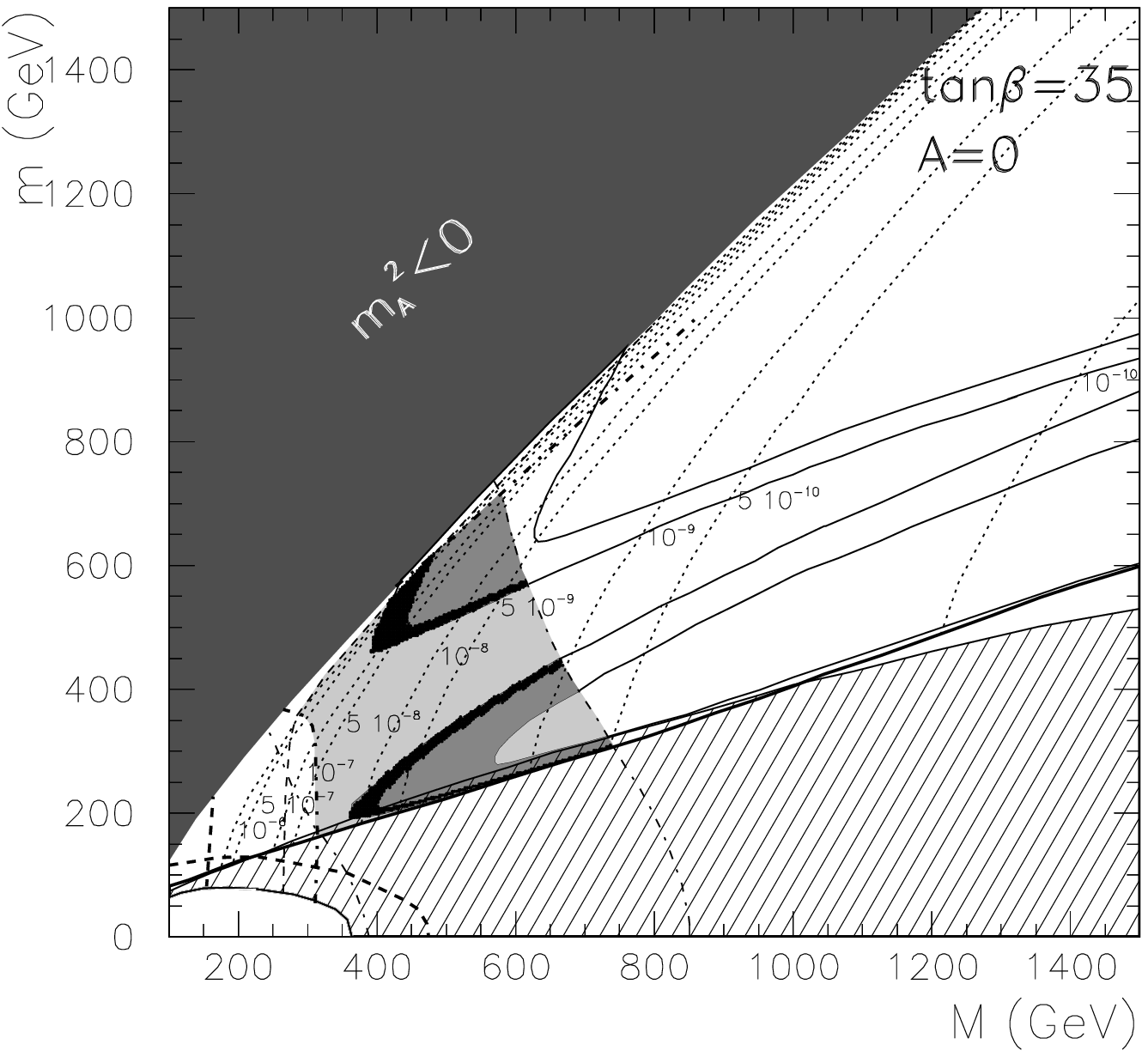,width=9cm}\,\,\,\,\,\,\,\,\,\,\,
%
\hspace*{-0.7cm}\epsfig{file=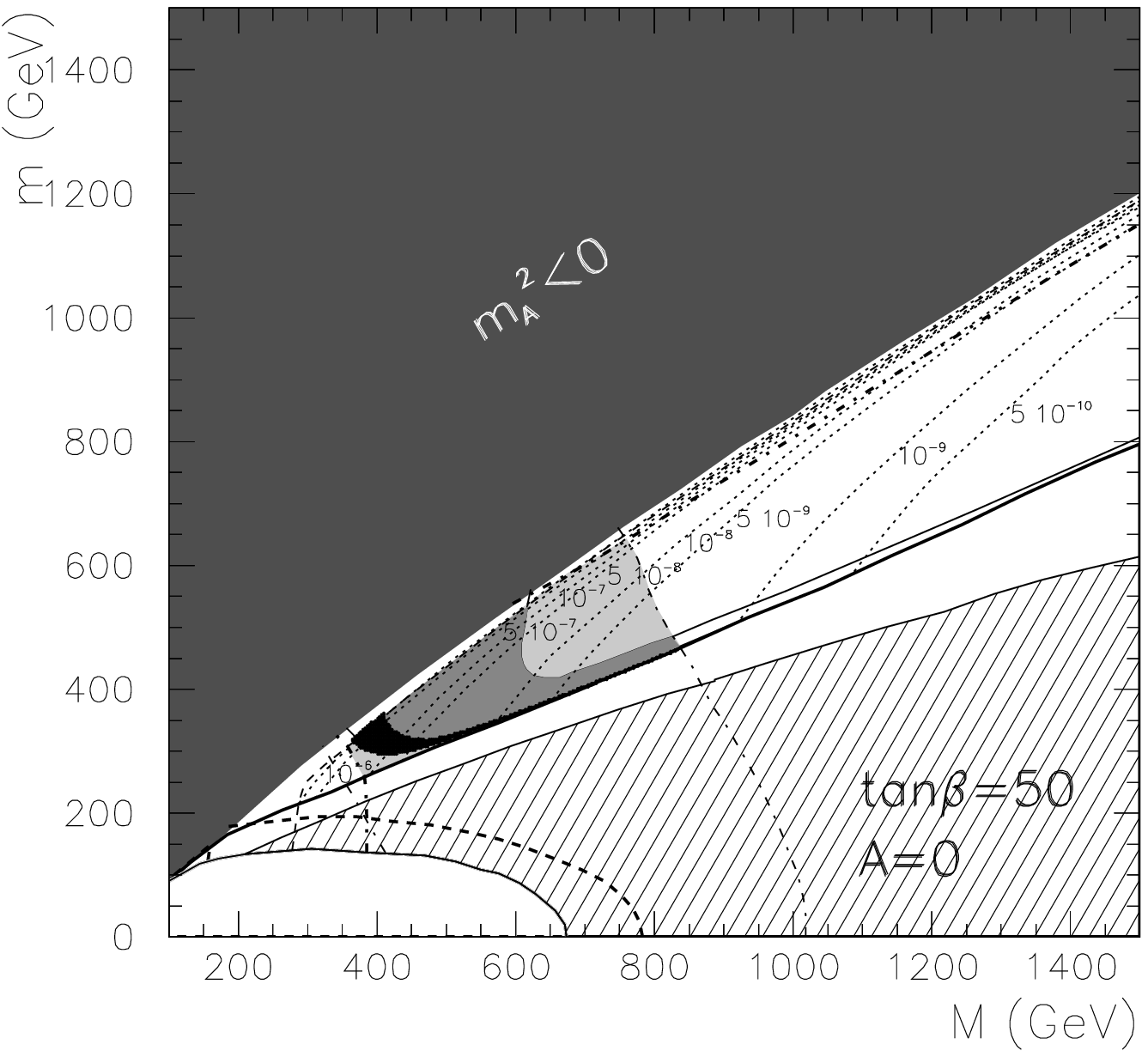,width=9cm} 



\captions{The same as in 
Figs.~2 and 3 
but for the non-universal case {\it c)} $\delta_1=-1$, $\delta_2=1$, 
discussed in eq.~(\ref{3cases}), with $\tgb=35,50$ and $A=0$.
\label{anu3}}
\end{figure}

\begin{figure}
\hspace*{-1.2cm}\epsfig{file=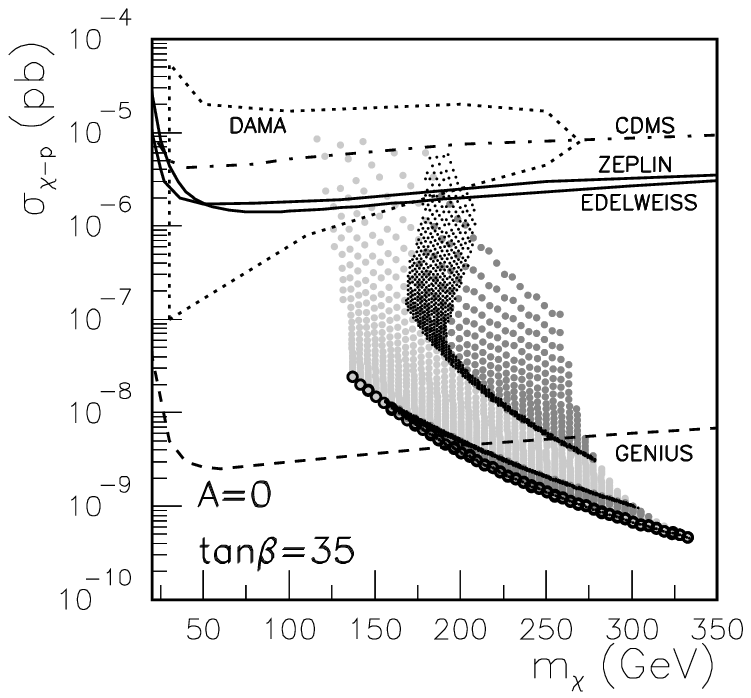,width=9cm}\,\,\,\,\,\,\,\,\,\,\,
%
\hspace*{-1.2cm}\epsfig{file=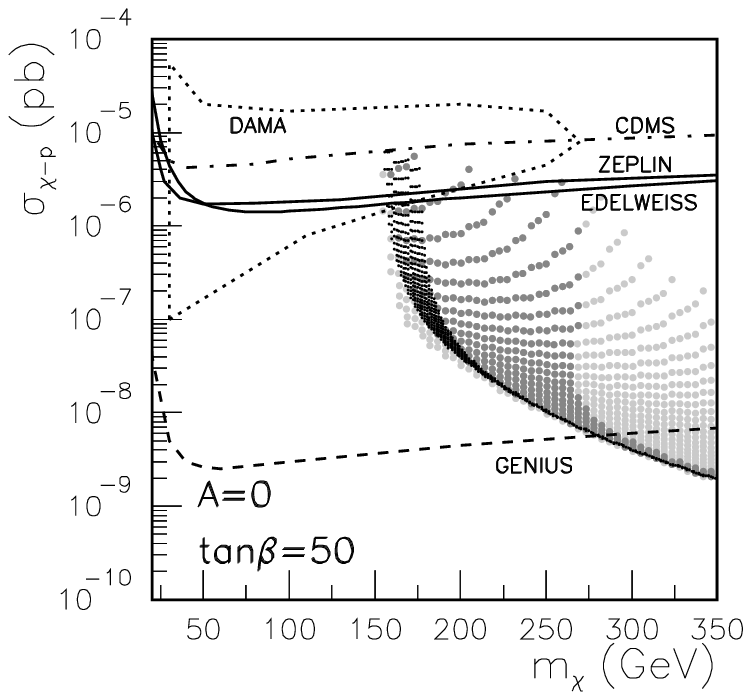,width=9cm}


\captions{The same as in Figs.~\ref{cross_scale122} and 5
but for the non-universal case {\it c)} $\delta_1=-1$, $\delta_2=1$, 
discussed in eq.~(\ref{3cases}), with $\tgb=35,50$ and $A=0$.
\label{anusec3}}
\end{figure}

\begin{figure}
\begin{center}
\epsfig{file=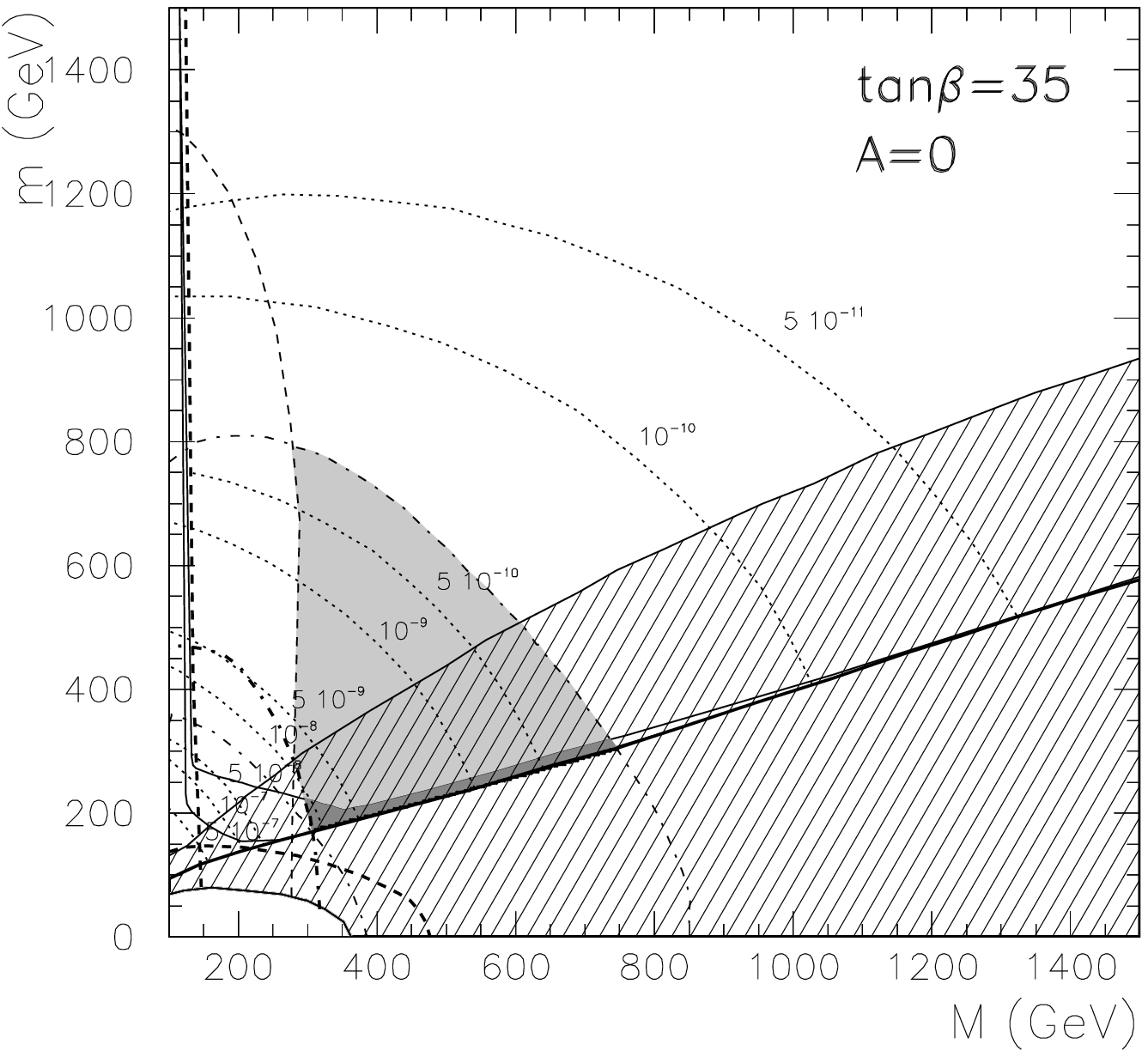,width=9cm}
\end{center}
%

%

\captions{The same as in Fig.~12 
but for $\delta_1=0$, $\delta_2=-1$, with $\tan\beta=35$ and
$A=0$.
\label{nuniversal}}
\end{figure}

Let us finally remark that although $\sin^2(\alpha-\beta)$ is close to 1 in
most of the points, a small number of them can have smaller values
when $\tan\beta=50$. These are points close to the region with
$m_A^2<0$, and therefore with small values for $m_A$. The same
situation occurs for the other cases studied below. 
Thus, according to our discussion
in Section~3, we use for these points the appropriate bound on the Higgs
mass \cite{barate}. In particular, in Fig.~13, the light grey dots
above the CDMS line correspond to these points.

For case {\it b)} the cross section increases also substantially
with respect to the universal case.
Now $\delta_2$ is taken vanishing and therefore the value 
of $\mu$ is essentially not modified with respect to the universal
case.
However, taking $\delta_1=-1$
produces an increase in the
cross section through 
the decrease in $m_A^2$,
as discussed previously. 
Note that for this case 
there is an area in the upper left of Fig.~14
where $m_A^2$ becomes negative. As
shown explicitly in Fig.~15, for
$\tan\beta=35$ and $A=0$,
there are points in the DAMA region. 
All of them correspond to $\sin^2(\alpha-\beta)$ not close to 1, and therefore
with an experimental bound on the Higgs mass smaller than 114.1 GeV.
All points with $\sin^2(\alpha-\beta)\sim 1$ have $\sigma_{\tilde{\chi}_1^0-p}\lsim 6\times 10^{-7}$ pb.
Note that points
with large values of the cross section
fulfil in this case the astrophysical bound
$\Omega_{\tilde{\chi}_1^0}h^2\gsim 0.1$.
Large values of $m$ reduce the resonant effects produced
by the Higgs $A$, and are sufficient to place the relic
abundance inside the bounds.
For $\tan\beta=50$, similarly to case {\it a)}, there are points
entering in the DAMA 
area, and part of them
fulfil the astrophysical bounds. Those above the ZEPLIN line correspond
to $\sin^2(\alpha-\beta)$ not close to 1.
For $A=M$ the figures as similar.

We have checked that smaller values of $\delta_1$,
as e.g. $\delta_1=-1.5,-2$, give also rise to

\begin{figure}
\hspace*{-1.2cm}\epsfig{file=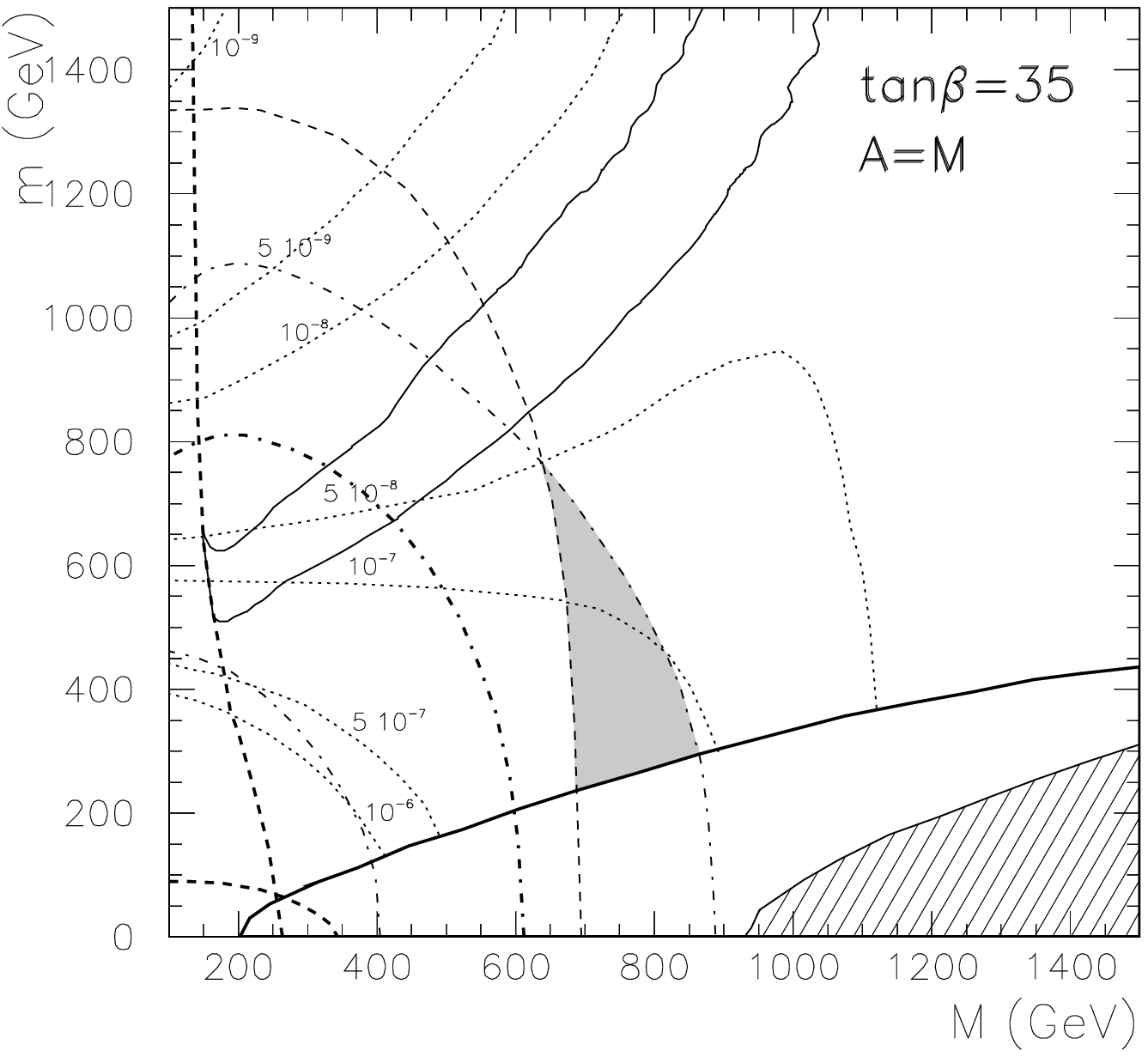,width=9cm}\,\,\,\,\,\,\,\,\,\,\,
\hspace*{-1.2cm}\epsfig{file=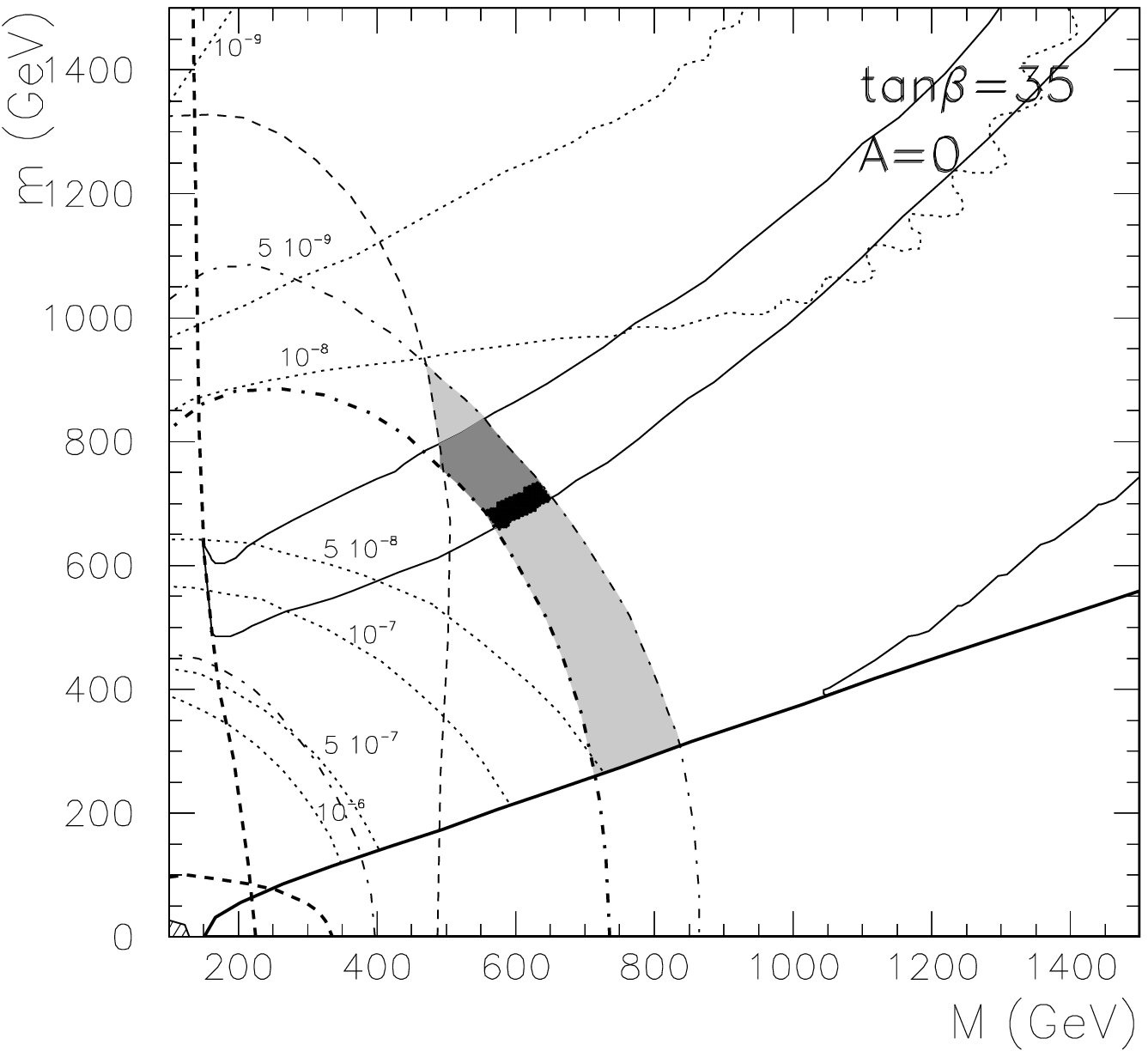,width=9cm}

\captions{
The same as in Fig.~2
but for 
the case discussed in eq.~(\ref{gauginospara}) with
non-universal soft gaugino masses, 
$\delta'_{1,2}=0, \delta'_3=-0.5$, taking
$\tan\beta=35$ and $A=M,0$.
\label{nunivg.eps}}
\end{figure}

\begin{figure}
\begin{center}
\epsfig{file=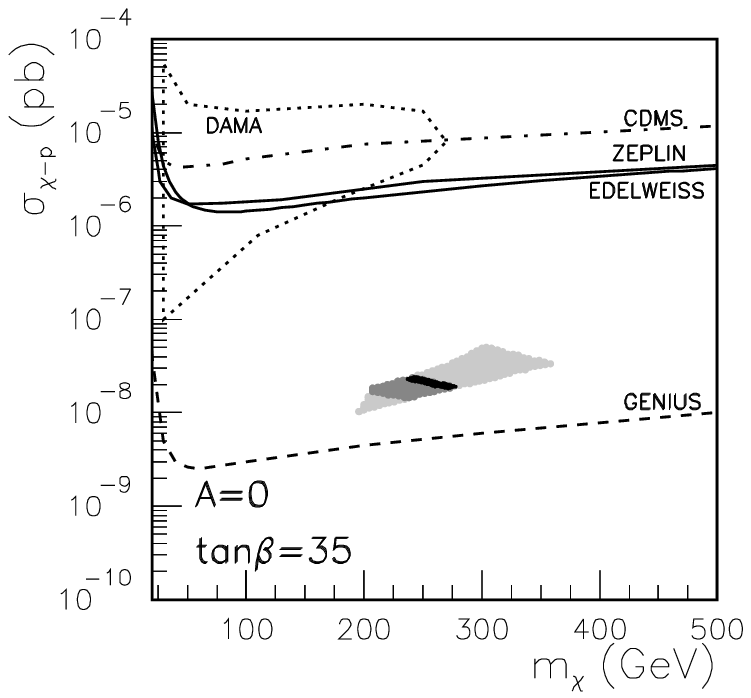,width=9cm}
\end{center}

\captions{The same as in Fig.~4
but for 
the case discussed in eq.~(\ref{gauginospara}) with
non-universal soft gaugino masses, 
$\delta'_{1,2}=0, \delta'_3=-0.5$, taking
$\tan\beta=35$ and $A=0$.
\label{nuniversal2}}
\end{figure}

\begin{figure}
\begin{center}
\hspace*{1.5cm}\epsfig{file=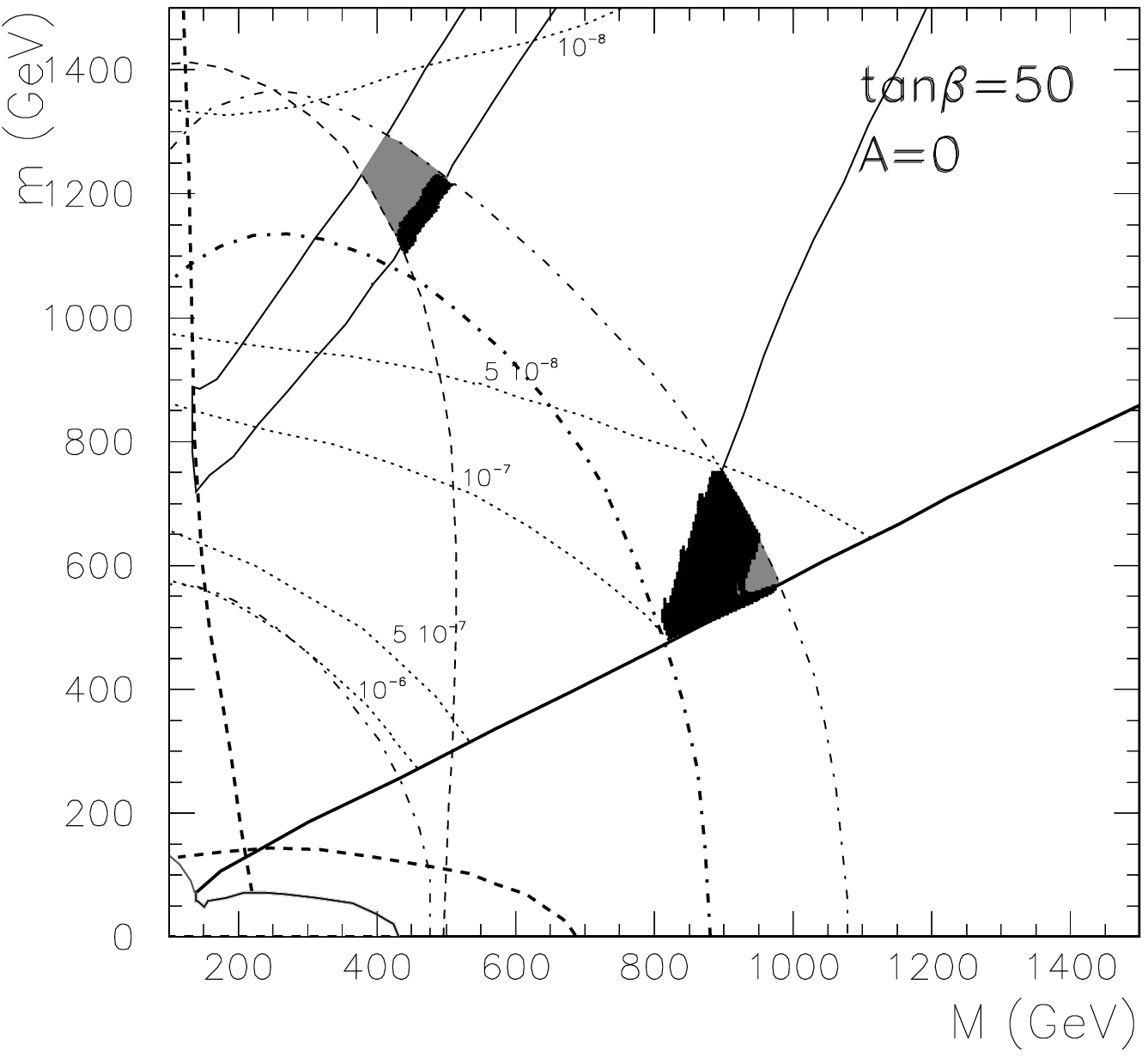,width=9cm}
\end{center}

\captions{
The same as in Fig.~19
but for 
$\tan\beta=50$ and $A=0$.
\label{nunivg4.eps}}
\end{figure}

\begin{figure}
\begin{center}
\epsfig{file=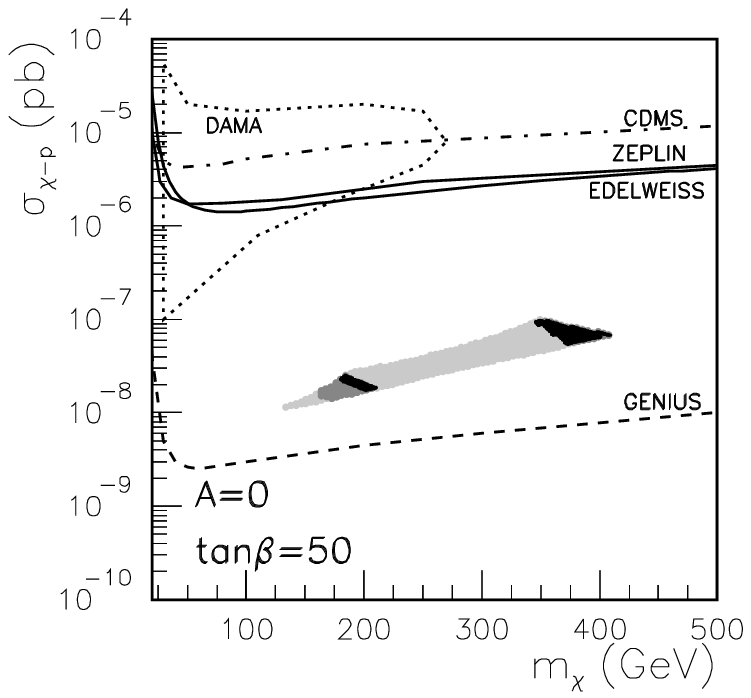,width=9cm}
\end{center}

\captions{The same as in Fig.~20
but for 
$\tan\beta=50$ and $A=0$.
\label{nuniversal23}}
\end{figure}

\noindent similar figures.
For larger values, $\delta_1\lsim -0.4$ is sufficient to enter in DAMA
fulfilling the
experimental and astrophysical bounds
with $\tan\beta=50$.

Finally, given the above situation concerning the enhancement
of the neutralino-proton cross section for {\it a)} and 
{\it b)}, it is clear that the combination of both cases
might be interesting.
This is carried out
in case {\it c)} where we take $\delta_1=-1$ and 
$\delta_2=1$. As shown in Figs.~16 and 17, cross sections 
as large as $\sigma_{\tilde{\chi}_1^0-p}\gsim 10^{-6}$ pb, entering
in DAMA
and fulfilling all experimental and astrophysical bounds, 
can be obtained for $\tan\beta=35,50$ and $A=0$. 
Those above the ZEPLIN line correspond
to $\sin^2(\alpha-\beta)$ not close to 1.
On the other hand, for $A=M$ and $\tan\beta=35$ no points with the
correct relic density enter in DAMA.
For other cases the results are similar.
For example, if we consider 
$\delta_1=-0.5$ and $\delta_2=1$, one obtains for
$\tan\beta=35$ points entering in DAMA but with
$\Omega_{\tilde{\chi}_1^0}h^2< 0.01$. For 
$\tan\beta=50$ a similar plot to the one in
Fig.~17 is obtained.

Concerning the restrictions coming from the UFB-3 constraint, 
we can see in Figs.~12, 14 and 16 
that these are slightly less important than in the
universal scenario (see Figs.~2 and 3).
For example, for case {\it a)}, as 
in the intermediate scale case, the main reason is that
$m_{H_u}^2$ at low energy becomes less negative. Thus the value of
$\ufbiii$ is increased, making the UFB-3 constraint less restrictive. 
For comparison, we 
show in Fig.~18
the same case as in
Fig.~12, for $\tan\beta=35$, 
but using the opposite choice for the sign of the
$\delta$ parameters.
Not only the cross section is smaller,
$\sigma_{\tilde{\chi}_1^0-p}< 10^{-8}$ pb,
but also the UFB-3 constraint is very
restrictive, forbidding all points which are allowed by
the experimental and astrophysical constraints.

\vspace{0.6cm}

\noindent {\it (ii) Non-universal gaugino masses}

\vspace{0.5cm}

Concerning gaugino masses,
let us parameterize their non-universality at $M_{GUT}$
as follows:
\begin{eqnarray}
M_1=M(1+\delta'_{1})\ , \quad M_2=M(1+ \delta'_{2})\ ,
\quad M_3=M(1+ \delta'_{3})
\ ,
\label{gauginospara}
\end{eqnarray}
where $M_{1,2,3}$ are the bino, wino and gluino masses, respectively.
Let us discuss now which values of the parameters are interesting in
order to increase the cross section with respect to the universal
case $\delta'_i=0$. In this sense,
it is worth noticing that
$M_3$ appears in the RGEs of squark masses, so e.g.
their contribution proportional
to the top Yukawa coupling in the RGE of $m_{H_u}^2$ will
do this less negative if $M_3$ is small, and therefore $\mu^2$ will become 
smaller in this case. 
However, small values of $M_3$ also lead to
an important decrease in the Higgs 
mass\footnote{We have checked that generically the
value of $\sin^2(\alpha-\beta)$ is very close to 1, and therefore
we are using the lower bound $m_h=114.1$ GeV as in the mSUGRA scenario.}.
In addition,
$b \to s \gamma$ and $g_{\mu}-2$ constraints are also relevant.
We show this in Fig.~19
for $\tan\beta=35$ and $A=M,0$, using
$\delta'_{1,2}=0, \delta'_3=-0.5$. 
As can be seen in the figure,
although the cross section increases with respect to the universal
case, the present experimental constraints exclude points
entering in the DAMA region.
On the other hand, only for $A=0$ there are
points allowed by all experimental and astrophysical constraints, and
they correspond to 
$\sigma_{\tilde{\chi}_1^0-p}\approx 3\times 10^{-8}$ pb.
Note that in this case $\Omega_{\tilde{\chi}_1^0}h^2$ 
decreases with respect to the universal case 
due to the enhancement of wino and
Higgsino components of the  ${\tilde{\chi}_1^0}$. The corresponding
increase of the annihilation cross section makes possible the existence of the
bands displayed in the figure without appealing to any kind of
coannihilations.
The result concerning the cross section is summarized
in Fig.~20.
Other values of $\delta'$s lead to qualitatively similar results.
For $\tan\beta=50$ the situation is similar, and we show the case $A=0$
in Figs.~21 and 22.

Finally, as in the previous case with non-universal scalars,
increasing the cross section through values at low energy
of $m_{H_u}^2$ less negatives implies
less important UFB constraints.
Now these are not very relevant, and in fact they correspond
to the UFB-1 ones.

\section{Conclusions}

We have carried out a theoretical analysis of the possibility
of detecting dark matter directly in current and projected
experiments. In particular, we have studied the value of the 
neutralino-nucleon
cross section in several supergravity scenarios.
In addition to the usual experimental and astrophysical constraints
we have imposed on the parameter space 
the absence of dangerous charge
and colour breaking minima.
This constraint, in particular the UFB-3, 
turns out to be quite important in some cases.
For example, in the usual mSUGRA scenario, where the soft terms
are assumed to be universal, and the GUT scale is considered, 
$\tan\beta\lsim 20$ is forbidden on this ground. In fact, even
larger values of $\tan\beta$ can also be forbidden, depending
on the value of the trilinear parameter $A$.
Concerning the cross section, this is
constrained to be 
$\sigma_{\tilde{\chi}_1^0-p}\lsim 3\times 10^{-8}$ pb, 
and therefore below the DAMA-reported data. Obviously, more sensitive
detectors, as e.g. GENIUS where values
of the cross section as low as $10^{-9}$ pb will be accessible, are needed.

When an intermediate scale is considered,
the running of the parameters is shorter, and in particular
$m_{H_u}^2$ is less negative at low energy 
producing a decrease in the value of  
$\mu^2$.
Although this effect increases the cross section, it
is not sufficient yet to enter in the DAMA area because of the
experimental bounds, 
which impose
$\sigma_{\tilde{\chi}_1^0-p}\lsim 4\times 10^{-7}$ pb.
In fact, at the end of the day, the preferred astrophysical range
for the relic neutralino density, 
$0.1\leq\Omega_{\tilde{\chi}_1^0}h^2\leq 0.3$,
imposes 
$\sigma_{\tilde{\chi}_1^0-p}\lsim  10^{-7}$ pb.
On the other hand, when $m_{H_u}^2$ is less negative
the negative contribution to 
$V_{UFB-3}$ is less important, and therefore also is 
less important the UFB-3 constraint. In fact, values of $\tan\beta$
smaller than 20 are now allowed. 

A similar situation concerning the UFB-3 constraint occurs when
non-universal soft terms are allowed and we try to increase
the cross section, since this is carried out by imposing 
a less negative value of $m_{H_u}^2$ at low energy. 
As a consequence the UFB-3 constraint is slightly less
important than in the universal scenario.
The other possibility to achieve this, to decrease 
$m_{H_d}^2$, does not modify essentially the UFB-3 constraint.
Of course, if the opposite procedure
is carried out, choosing the
parameters in such a way that the cross section decreases,
the UFB-3 constraint is more important.

Concerning the cross section, this can 
be increased a lot with respect to the 
universal scenario, when non-universal scalars are considered. 
It is even possible, for some particular values of the
parameters, to find points entering
in DAMA,
$\sigma_{\tilde{\chi}_1^0-p}\approx 10^{-6}$ pb,
and fulfilling all experimental and astrophysical constraints.
On the contrary, when non-universal gauginos are considered,
although the cross section increases, the experimental bounds
exclude the possibility of entering in DAMA. 




\vspace{1cm}

\noindent {\bf Acknowledgements}

\noindent
The work of D.G. Cerde\~no 
was supported by the Deutsche Forschungsgemeinschaft. 
D.G. Cerde\~no also thanks the MLU
Halle-Wittenberg for their hospitality during the early stages of this
work.
E. Gabrielli would like to thank the Academy of Finland 
(project number 48787) for financial support.
M.E. Gomez acknowledges support from the Funda\c
c\~ao para a Ci\^encia e Tecnologia under contract
SFRH/BPD/5711/2001 and project CFIF-Plurianual (2/91).
The work of C. Mu\~noz was supported in part by the Ministerio de 
Ciencia y Tecnolog\'{\i}a under contract FPA2000-0980, 
and the European Union under contract 
HPRN-CT-2000-00148.

\end{document}